\documentclass{elsart}
\usepackage{graphics}

\begin{document}

\def\etal{{\it{}et~al.}}        
\def\eref#1{(\protect\ref{#1})}
\def\fref#1{\protect\ref{#1}}
\def\sref#1{\protect\ref{#1}}
\def\tref#1{\protect\ref{#1}}
\def\av#1{\langle#1\rangle}
\def\th#1{$#1^{\rm th}$}

%
\def\tcapt#1{\refstepcounter{table}\bigskip\hbox to \textwidth{%
       \hfil\vbox{\hsize=\captwidth\renewcommand{\baselinestretch}{1}\small
       {\sc Table \thetable}\quad#1}\hfil}\bigskip}

%
\newdimen\captwidth
\captwidth=13cm                                
\newdimen\sidecaptwidth
\sidecaptwidth=5cm                             
\newdimen\normalfigwidth
\normalfigwidth=\captwidth                     
\newdimen\smallfigwidth
\smallfigwidth=11cm                            
\newdimen\tinyfigwidth
\tinyfigwidth=4cm                              
\newdimen\realtinyfigwidth
\realtinyfigwidth=3cm                          
\newdimen\sidefigwidth
\sidefigwidth=7cm                              

\def\capt#1{\refstepcounter{figure}\bigskip\hbox to \textwidth{%
       \hfil\vbox{\hsize=\captwidth\renewcommand{\baselinestretch}{1}\small
       {\sc Figure \thefigure}\quad#1}\hfil}\bigskip}

\def\normalfigure#1{\hbox to\textwidth{%
       \hfil\resizebox{\normalfigwidth}{!}{\includegraphics{#1}}\hfil}}

\def\smallfigure#1{\hbox to\textwidth{%
       \hfil\resizebox{\smallfigwidth}{!}{\includegraphics{#1}}\hfil}}

\def\ninefigure#1#2#3#4#5#6#7#8#9{\hbox to \textwidth{%
       \hfil\hbox to\normalfigwidth{%
       \resizebox{\tinyfigwidth}{!}{\includegraphics{#1}}\hfil
       \resizebox{\tinyfigwidth}{!}{\includegraphics{#2}}\hfil
       \resizebox{\tinyfigwidth}{!}{\includegraphics{#3}}}\hfil}
       \bigskip
       \hbox to \textwidth{%
       \hfil\hbox to\normalfigwidth{%
       \resizebox{\tinyfigwidth}{!}{\includegraphics{#4}}\hfil
       \resizebox{\tinyfigwidth}{!}{\includegraphics{#5}}\hfil
       \resizebox{\tinyfigwidth}{!}{\includegraphics{#6}}}\hfil}
       \bigskip
       \hbox to \textwidth{%
       \hfil\hbox to\normalfigwidth{%
       \resizebox{\tinyfigwidth}{!}{\includegraphics{#7}}\hfil
       \resizebox{\tinyfigwidth}{!}{\includegraphics{#8}}\hfil
       \resizebox{\tinyfigwidth}{!}{\includegraphics{#9}}}\hfil}}

\def\fourfigure#1#2#3#4{\hbox to \textwidth{%
       \hfil\hbox to\normalfigwidth{%
       \resizebox{\realtinyfigwidth}{!}{\includegraphics{#1}}\hfil
       \resizebox{\realtinyfigwidth}{!}{\includegraphics{#2}}\hfil
       \resizebox{\realtinyfigwidth}{!}{\includegraphics{#3}}\hfil
       \resizebox{\realtinyfigwidth}{!}{\includegraphics{#4}}}\hfil}}

\def\sidefigure#1#2{\hbox to \textwidth{%
       \resizebox{\sidefigwidth}{!}{\includegraphics{#1}}\hfil
       \refstepcounter{figure}
       \vbox{\hsize=\sidecaptwidth\renewcommand{\baselinestretch}{1}
       \small\raggedright{\sc Figure \thefigure}\quad#2}}}

\date{16 June 1997}
\journal{Physical Review E}

\begin{frontmatter}
\title{Monte Carlo simulation of ice models}
\author{G. T. Barkema}
\address{HLRZ, Forschungszentrum J\"ulich, 52425 J\"ulich, Germany}
\author{M. E. J. Newman}
\address{Santa Fe Institute, 1399 Hyde Park Road, Santa Fe, NM 87501.  U.S.A.}
\begin{abstract}
  We propose a number of Monte Carlo algorithms for the simulation of ice
  models and compare their efficiency.  One of them, a cluster algorithm
  for the equivalent three colour model, appears to have a dynamic exponent
  close to zero, making it particularly useful for simulations of critical
  ice models.  We have performed extensive simulations using our algorithms
  to determine a number of critical exponents for the square ice and F
  models.
\end{abstract}
\end{frontmatter}

\section{Introduction}
\label{intro}
Ice models are a class of simple classical models of the statistical
properties of the hydrogen atoms in water ice.  In ice, the oxygen atoms
are located on a lattice, and each oxygen atom has four hydrogen bonds to
neighbouring oxygen atoms, giving a four-fold coordinated lattice.
However, as has long been known, the proton (hydrogen atom) which forms a
hydrogen bond is located not at the centre point of the line between two
oxygens, but at a point closer to one of the two.  Bernal and
Fowler~\cite{Bernal33} and Pauling~\cite{Pauling35} proposed that the
protons are arranged according to two rules, known as the ice rules:

\begin{enumerate}
\item There is precisely one hydrogen atom on each hydrogen bond.
\item There are precisely two hydrogen atoms near each oxygen atom.
\end{enumerate}

Ice models are a class of models mimicking the behaviour of systems which
obey these rules.  The most widely-studied ice model is the model on a
square lattice in two dimensions.  A version of this model has been solved
exactly by Lieb~\cite{Lieb67a,Lieb67b,Baxter82}.  The exact solution gives
us, for instance, the critical temperature and the free energy of the
model.  However, there are a number of quantities of interest which cannot
be obtained from the exact solution, and for these quantities we turn to
Monte Carlo simulation.

In this paper we introduce a number of new Monte Carlo algorithms for the
simulation of ice models, and compare their efficiency.  We will show that
one of them, the three-colour cluster algorithm developed in
Section~\sref{mctc}, possesses a very small dynamic exponent (possibly
zero), and so suffers very little from critical slowing down.  Using this,
and other algorithms presented here, we determine numerically several
critical exponents which have not been accurately measured previously: the
dimensionality of the percolating cluster of symmetric vertices in the F
model at critical temperature, the scaling of the largest loop in the
loop-representation of both square ice and the F model at critical
temperature, and the scaling of the trajectory of a wandering defect in
square ice.

\section{Ice models}
\label{icemodels}
\begin{figure}
\normalfigure{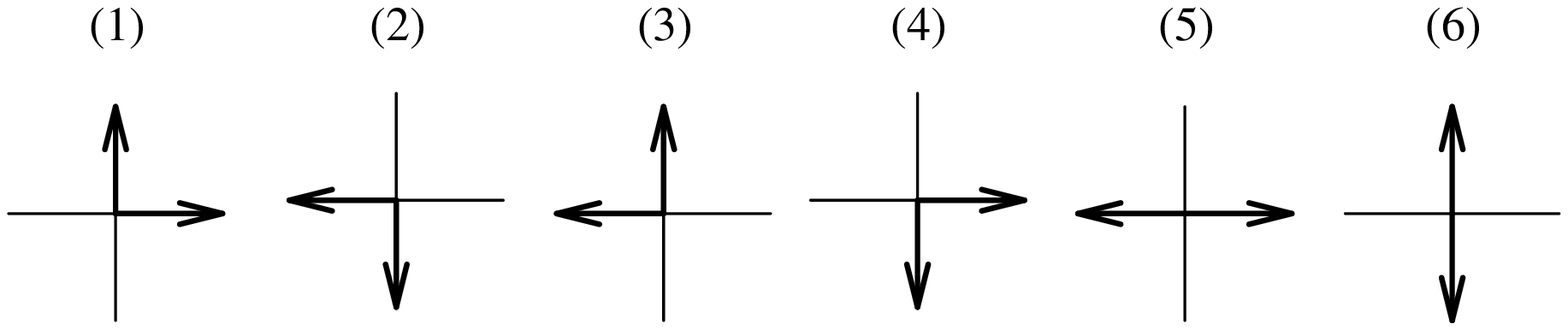}
\capt{The six possible vertex configurations of an ice model on a
  square lattice.}
\label{sixvertex}
\end{figure}

Our ice model is as follows.  Oxygen atoms are arranged on the vertices of
a square grid, and between each oxygen and its four neighbours there are
hydrogen bonds, represented by the lines of the grid.  Commonly, arrows are
drawn on the bonds to indicate the positions of the protons: the arrow
points towards the vertex which the proton is nearest to.  The first ice
rule above then corresponds to the condition that there should be exactly
one arrow on each bond.  The second ice rule says that each vertex should
have exactly two arrows pointing towards it, and two pointing away.  This
gives us six types of vertices, and for this reason ice models are
sometimes also referred to as six-vertex models.  The six vertices are
illustrated in Figure~\fref{sixvertex}.

In the first part of this paper we study the simplest six-vertex model, in
which all types of vertices are assigned the same energy.  This model is
usually called ``square ice''.  The name is somewhat confusing, since other
ice models on square lattices, such as the KDP and F models of
Section~\sref{energies}, are not also called square ice.  However, since
the name is widely used we will follow convention and use it here too.
Because all configurations of square ice possess the same energy, the
model's properties are entirely entropically driven and variations in
temperature have no effect on its behaviour.

It turns out that the square ice model is equivalent to two other
well-studied models in statistical physics: the three-colouring model, and
a random-surface model on a square lattice.  In this section we describe
these two models as well as introducing the ``fully loop-covered lattice''
model, which is also equivalent to the square ice model.

\subsection{Colouring models}
\label{threecolouring}
\begin{figure}
\sidefigure{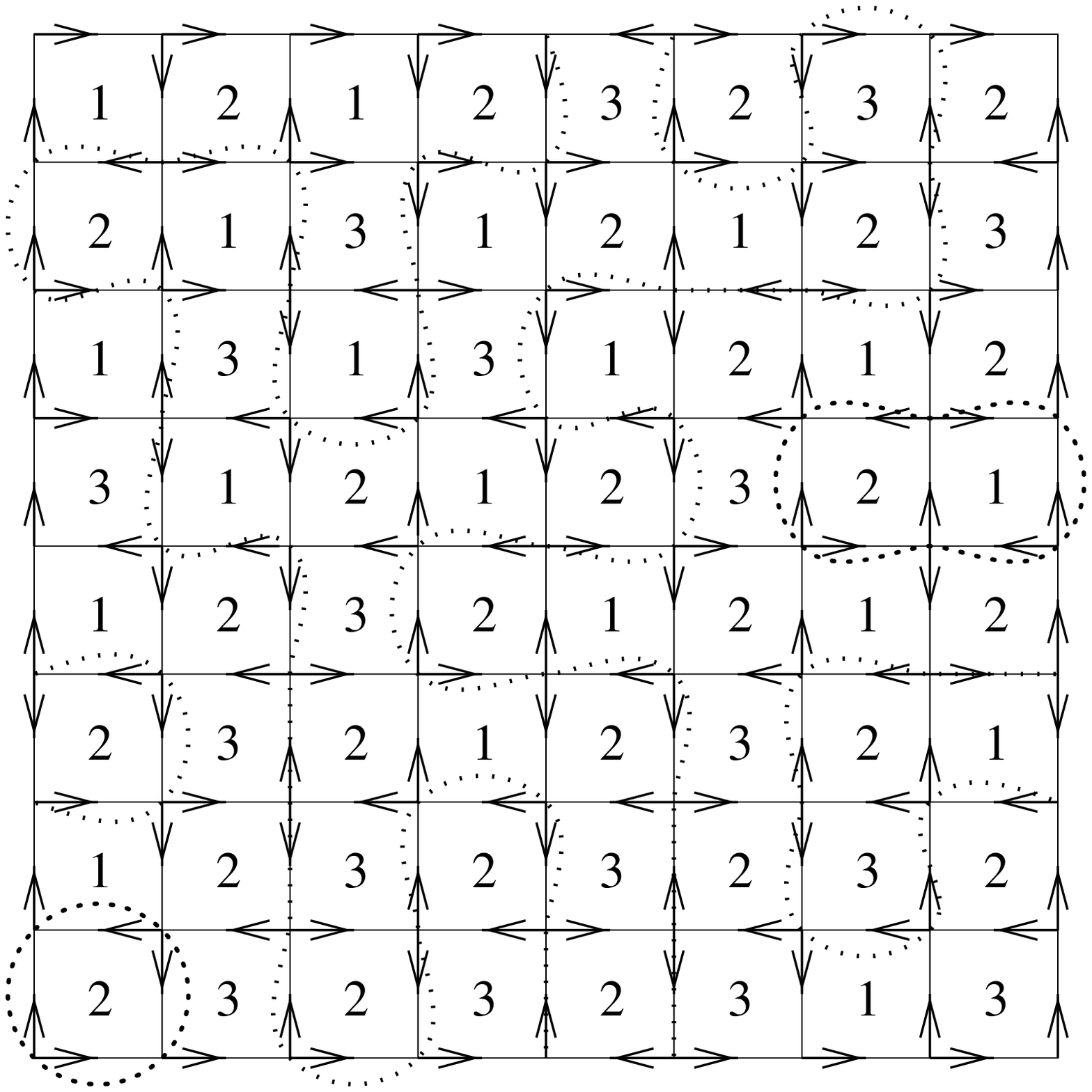}{A three-colouring of a square lattice, its
  corresponding configuration of arrows, and its corresponding
  loop-covering.}
\label{equivalent}
\end{figure}

Lenard~\cite{Lenard,Lieb67a} has shown an important result about square ice
which will help us in the design of an efficient Monte Carlo algorithm for
the simulation of the model.  Lenard demonstrated that the configurations
of an ice model on a square lattice can be mapped onto the configurations
of a lattice of squares coloured using three different colours, with the
restriction that no two nearest-neighbour squares have the same colour.  It
is actually not very difficult to demonstrate this equivalence.  The
procedure for working out the configuration of the arrows of the ice model,
given a suitable colouring of the plaquets on the lattice is shown in
Figure~\fref{equivalent}, in which the three colours are represented by the
numbers 1, 2 and 3.  The rule is that we imagine ourselves standing on one
of the squares of the lattice and looking towards one of the adjacent ones.
If the number in the adjacent square is one higher (modulo three) than the
number in the square we are standing on, we draw an arrow on the
intervening bond pointing to the right.  Otherwise we draw an arrow to the
left.  The procedure is then repeated for every other bond on the lattice.

Clearly the resulting configuration of arrows obeys the first ice rule;
since neighbouring plaquets must have different colours the prescription
above will place one and only one arrow on every bond.  The second ice rule
requires that each vertex has two ingoing and two outgoing arrows.  If we
walk from square to square in four steps around a vertex, then each time we
cross a bond separating two squares, the colour either increases or
decreases by one, modulo three.  The only way to get back to the colour we
started with when we have gone all the way around is if we increase twice
and decrease twice.  This means that the vertex we walk around must have
two ingoing and two outgoing arrows, exactly as we desire.  Thus each
configuration of the three-colouring model corresponds to a unique correct
configuration of square ice.

We can also reverse the process, transforming an ice model configuration
into a three-colouring.  We are free to choose the colour of one square on
the lattice as we wish, but once that one is fixed, the arrows on the bonds
separating that square from each of its neighbours uniquely determine the
colour of the neighbouring squares, and, by repeated application of the
rule given above, the colour of all the rest of the squares in the lattice.
Thus, the number of ways in which the squares of the lattice can be
coloured is exactly the number of configurations of the ice model on the
same lattice, regardless of the size of the lattice, except for a factor of
three.

\subsection{Random surfaces}
\label{randomsurfaces}
Square ice is also equivalent to a random surface model in which heights
are assigned to the plaquets of a square lattice.  If we assign these
heights in such a way that adjacent plaquets have heights which differ by
exactly 1, then again there is a one-to-one mapping between the
configurations of the ice model and the random surface: the mapping is
identical to the three colour mapping of the last section except for the
absence of the modulo operation.

\subsection{Fully loop-covered lattices}
\label{fullloop}
The six-vertex model is also equivalent to a ``fully loop-covered lattice
model'' in which (non-directed) loops are formed by joining the vertices of
the square lattice with ``links'' in such a way that each site on the
lattice belongs to exactly one self-avoiding loop.  To demonstrate this
equivalence, consider the following rule.  First, divide the lattice into
even and odd sites in a checkerboard pattern.  Now place links along all
bonds whose arrows are pointing away from an even vertex.  Since each such
arrow must also be pointing towards an odd vertex, and since each vertex
has two ingoing and two outgoing arrows, this creates two links to every
site on the lattice.  Hence the lattice is fully covered by closed
self-avoiding loops.

Proving the reverse result, that each configuration of loops corresponds to
exactly one configuration of arrows is equally simple: we place outgoing
arrows on each bond adjoining an even site which is part of one of our
loops.  The direction of all the remaining arrows is then fixed by using
the second ice rule.

\section{Monte Carlo algorithms for square ice}
\label{longloops}
In this paper we develop a number of different Monte Carlo algorithms for
calculating the average properties of ice models on square lattices.  In
the case of square ice, in which all configurations of the lattice have the
same energy, the necessary steps for creating such an algorithm are (i)~to
choose a set of elementary moves which take us from one state of the model
to another, (ii)~to demonstrate that these moves can take us from any state
of a finite lattice to any other in a finite number of steps (the condition
of ergodicity) and (iii)~to construct an algorithm from these moves such
that in equilibrium the rate at which a particular move occurs which takes
us from a state $\mu$ to a state $\nu$ is the same as the rate for the
reverse move from $\nu$ to $\mu$ (the condition of detailed balance).  It
is then straightforward to show that over a sufficiently long period of
time we will sample all states on a finite lattice with equal probability.
The choice of an elementary move however is not obvious, since there is no
local change we can make to the directions of the arrows on the lattice
which will preserve the second ice rule.  There is no equivalent of the
reversal of a single spin in an Ising model, for example.  In the next few
sections we will consider four different candidate non-local update moves
for square ice, which lead us to four different Monte Carlo algorithms of
varying efficiency.  First, we look at the standard algorithm, which
involves reversing the arrows around a loop on the lattice.

\subsection{The standard ice algorithm}
It is clear that one possible move which takes us from an allowed
configuration of arrows in an ice model to another is the reversal of all
the arrows in a loop chosen such that all arrows point in the same
direction around the loop.  Such a loop has one arrow pointing in and one
pointing out of each participating vertex, so that the reversal of all of
them preserves the number of arrows going in and out of each vertex.
Notice that these loops are not the same loops as those in the fully
loop-covered model described above.  In that case the arrows along the loop
point in alternating directions, and their reversal would violate the
second ice rule.

\begin{figure}
\normalfigure{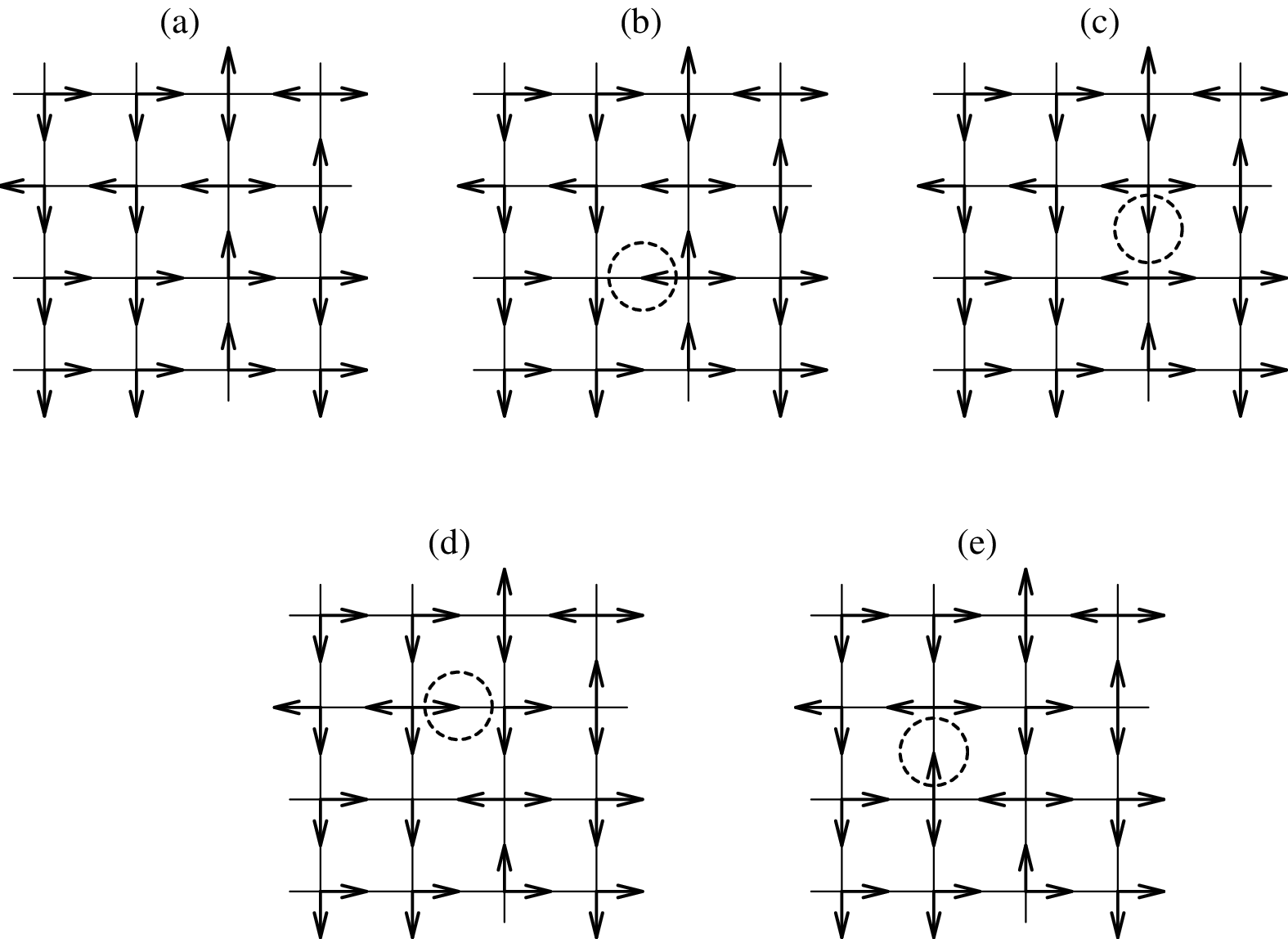}
\capt{Flipping arrows one by one along a line across the lattice allows us
to change the configuration and still satisfy the ice rules.  The only
problems are at the ends of the line, but if the two ends eventually
meet one another forming a closed loop of flipped arrows, this problem goes
away too.}
\label{ionic}
\end{figure}

How do we find a loop in which all arrows point in the same direction
around the loop?  The most straightforward method is illustrated in
Figure~\fref{ionic}.  Starting with a correct configuration of the model
(a), we choose a single vertex at random from the lattice.  We then choose
at random one of the two outgoing arrows at this vertex and reverse it (b).
(We could just as well choose an ingoing arrow---either is fine.)  This
creates a violation of the second ice rule at the initial vertex and also
at a new vertex at the other end of the reversed arrow.  In ice terminology
these are referred to as ionic defects: the vertices with one and three
outgoing arrows correspond to $OH^-$ and $H_3O^+$ respectively.  We can
remove the defect at the new vertex by choosing at random one of the two
outgoing arrows at this vertex and reversing it (c).  (There are actually
three outgoing arrows at this vertex, but one of them is the arrow we
reversed in the first move and we exclude this one from our set of possible
choices to avoid having the loop retrace its steps.)  This creates another
defect at the other end of that arrow, and so forth.  In this manner one of
the two defects created by the reversal of the first arrow diffuses around
the lattice (d) until by chance it finds itself back at the starting site
once more, at which point it annihilates with the defect there resulting in
a new configuration of the lattice which satisfies the ice rules (e).  The
net result is the reversal of a loop of arrows on the lattice.

In the figure we have illustrated the case of the smallest possible loop,
which on the square lattice involves the reversal of just four arrows.
However, provided the size of the lattice allows for it, the loops can be
arbitrarily long, and for this reason we will refer to this algorithm as
the ``long loop algorithm''.  At each step around the loop we have a choice
to make between two possible arrows that we could reverse, and if we make
these choices at random with equal probability we generate a species of
random walk across the lattice.  This walk could quite possibly take a long
time to return to its starting point.  However, on the finite lattices we
use in our Monte Carlo simulations we are guaranteed that the walk will
eventually return.  And long loops are not necessarily a bad thing, since
although they take longer to generate they also flip a larger number of
arrows, which allows the system to decorrelate quicker.

An alternative, but entirely equivalent scheme, makes use of so-called
Bjerrum defects~\cite{Bjerrum51}, rather than the ionic defects we have
employed.  A Bjerrum defect is a violation of the first ice rule: a bond
containing two protons, one at either end of the bond (a Bjerrum D defect),
or a bond containing no protons (a Bjerrum L defect).  One can construct a
Monte Carlo move using Bjerrum defects just as we did with ionic defects by
removing an arrow from a bond, and placing it on another bond.  This
creates one D defect and one L defect.  These defects can also wander
around and eventually recombine, resulting in a new state of the lattice.
Algorithms based on wandering Bjerrum defects have been used by Rahman and
Stillinger~\cite{Rahman72} for the simulation of three-dimensional ice and
by Yanagawa and Nagle~\cite{Yanagawa79} for the simulation of
two-dimensional ice.

The process in which two defects (either ionic or Bjerrum) are created and
diffuse around the lattice until they find one another again is actually
very similar to what goes on in real ice.  In real ice, changes in the
proton configuration are mediated principally by the diffusion of Bjerrum
defects around the lattice.  The density of defects is very small---already
at $-10^\circ$C only about one in five million bonds is occupied by a
Bjerrum defect---and the number of ionic defects is smaller even than
this~\cite{Hobbs74}.

\subsection{Ergodicity}
We have now specified a move that will take us from one correct
configuration of the arrows to another, and our proposed Monte Carlo
algorithm for square ice is simply to carry out a large number of such
moves, one after another.  However, as we remarked above, we still need to
demonstrate that the algorithm satisfies the criteria of ergodicity and
detailed balance.

\begin{figure}
\normalfigure{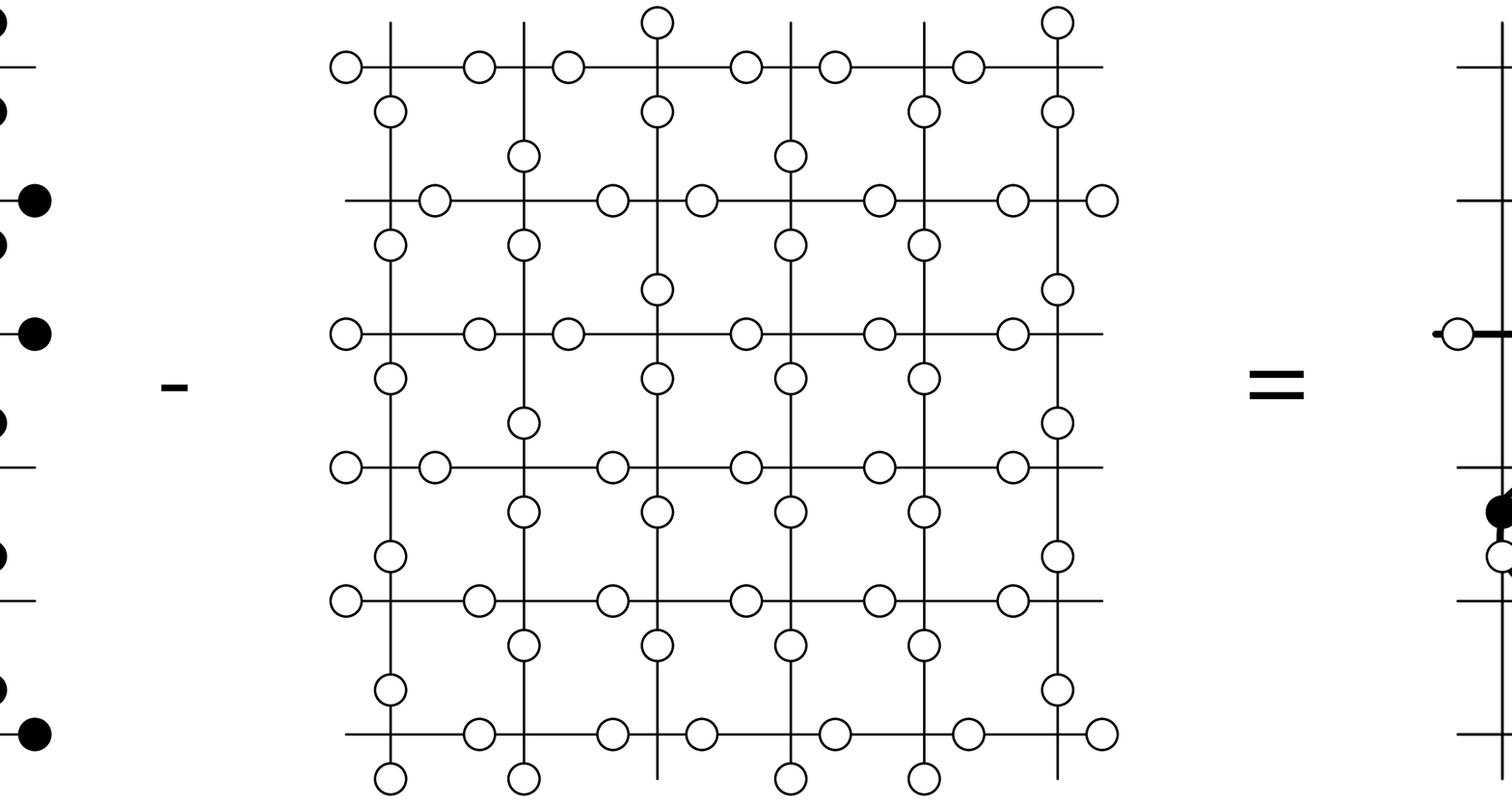}
\capt{The difference between any two configurations of the six-vertex model
  can be decomposed in a number of loops (which may run around the periodic
  boundaries).  If all the arrows along these loops are reversed, we go
  from one configuration to the other.}
\label{ergodicity}
\end{figure}

First, consider ergodicity, whose proof is illustrated in
Figure~\fref{ergodicity}.  The figure shows how the difference between two
configurations of the model on a finite lattice can be decomposed into the
flips of arrows around a finite number of loops.  We can demonstrate the
truth of this statement for any two configurations by the following
argument.  Each of the vertices in Figure~\fref{sixvertex} differs from
each of the others by the reversal of an even number of arrows.  This fact
follows directly from the ice rules.  Thus, if we take two different
configurations of the model on a particular lattice and imagine drawing
lines along the bonds on which the arrows differ, we are guaranteed that
there will be an even number of such lines meeting at each vertex.  Thus
these lines must form a set of (possibly intersecting) loops covering a
subset of the vertices on the lattice.  It is not difficult to show that
these loops can be chosen so that the arrows around each one all point in
the same direction.  Since the reversal of the arrows around these loops
are precisely our Monte Carlo moves, and since there are a finite number of
such loops, it follows that we can get from any configuration to any other
in a finite number of steps, and thus the system is ergodic.  Note that it
is important to allow the loops to pass through the periodic boundary
conditions for this to work.\footnote{It is not too hard to show that the
  loops which wrap around the periodic boundary conditions change the
  polarization, Equation~\eref{polarization}, of the system, whereas the
  ones which don't conserve polarization.  Thus, if we do not allow the
  loops to wrap around in this way the polarization would never change.}

\subsection{Detailed balance}
Our Monte Carlo move consists of choosing a starting site $S_0$ and
reversing a loop of arrows starting at that site and ending, $m$ steps
later, at the same site $S_m=S_0$.  The probability of selecting a
particular site $S_0$ as the starting site is $1/N$, where $N$ is the
number of sites on the lattice.  The probability of making a particular
choice from the two possible outgoing arrows at each step around the loop
is $\frac12$ for each step, so that the probability that we chose a certain
sequence of steps is equal to $2^{-m}$, and the probability of generating
the entire loop is $\frac1N 2^{-m}$.  For the reverse move, in which the
same loop of arrows is flipped back again to take us from state $\nu$ back
to state $\mu$, the exact same arguments apply, again giving us a
probability of $\frac1N 2^{-m}$ for making the move, and hence detailed
balance is observed.  This, in combination with the demonstration of
ergodicity above, ensures that our algorithm will sample all states of the
model with equal probability.

\section{An alternative algorithm, involving smaller loops}
\label{smallloops}
A practical problem which arises in the algorithm presented above, is that
if we simulate a large lattice, the probability that we return to the
starting site $S_0$ is quite small once we have wandered sufficiently far
away from it, and thus it may take a long time to generate even one move.
In response to this problem, we have devised a second algorithm which also
reverses the arrows around a closed loop of bonds, but this algorithm
generates shorter loops.  For obvious reasons we call this the ``short loop
algorithm''.  The short loop algorithm works in a similar way to the long
loop algorithm: we choose a starting site $S_0$ at random from the lattice
and reverse one of the outgoing arrows at that vertex, thereby creating two
defects.  We then reverse further arrows so that one of the defects wanders
around the lattice randomly.  However, rather than waiting until the two
defects find one another again, we now continue only until the wandering
defect encounters a site, call it $S_m$, which it has encountered before in
its path across the lattice: $S_m=S_l$ with $l<m$.  From this point, we
retrace our steps {\em backwards\/} down the old path of the defect, until
we reach $S_0$ again, reversing all the arrows along the way.  The net
result is that we reverse all the arrows along the path from site $S_0$ to
$S_l$ twice (which means that they are the same before and after the move),
and all the arrows in the loop from $S_l$ to $S_m$ once.  Thus we have
again reversed all the arrows around a loop.  By contrast with the long
loop algorithm however, the wandering defect does not have to find its way
back to its original starting point; it only needs to find any site on its
previous path.  This guarantees that the length of its walk will never
exceed $N$ steps, and in practice the typical move is much shorter than
this.  (In fact, the number of steps tends to a finite limit as the lattice
becomes large---see Section~\sref{shortsec}.)

As with the previous algorithm, we need to demonstrate ergodicity and
detailed balance.  The proof of ergodicity is identical to that for the
previous case: the difference between any two states on a finite lattice
can be reduced to the reversal of the spins around a finite number of
loops.  Since the algorithm has a finite chance of reversing each such
loop, it can connect any two states in a finite number of moves.

The proof of detailed balance is also similar to that for the long loop
algorithm.  Consider again a move which takes us from state $\mu$ to state
$\nu$.  The move consists of choosing a starting site $S_0$ at random, then
a path $P=\lbrace S_0 \ldots S_l \rbrace$ in which the arrows are left
untouched, followed by a loop $L=\lbrace S_l \ldots S_m \rbrace$ in which
we reverse the arrows.  (Remember that the last site in the loop $S_m$ is
necessarily the same as the first $S_l$.)  The probability that we chose
$S_0$ as the starting point is $1/N$, where $N$ is the number of sites on
the lattice.  After that we have a choice of two directions at each step
along the starting path and around the loop, so that the probability that
we end up taking the path $P$ is equal to $2^{-l}$ and the probability that
we follow the loop $L$ is $2^{-(m-l)}$.  After the loop reaches site
$S_m=S_l$, we do not have any more free choices.  The probability that we
move from a configuration $\mu$ to configuration $\nu$ by following a
particular path $P$ and loop $L$ is thus
\begin{equation}
P(\mu\to\nu) = \frac1N 2^{-l} 2^{-(m-l)} = 2^{-m}.
\label{path_forward}
\end{equation}
 
For the reverse move, the probability of starting at $S_0$ is again $1/N$,
and the probability of following the same path $P$ as before to site $S_l$
is $2^{-l}$ again.  However, we cannot now follow the same loop $L$ from
$S_l$ to $S_m$ as we did before, since the arrows along the loop are
reversed from what they were in state $\mu$.  On the other hand, we can
follow the loop in the reverse direction, and this again has probability
$2^{-(m-l)}$.  Thus we have
\begin{equation}
P(\nu\to\mu) = \frac1N 2^{-l} 2^{-(m-l)} = 2^{-m},
\label{path_backward}
\end{equation}
exactly as before.  This demonstrates detailed balance for the algorithm
and, in combination with the demonstration of ergodicity, ensures that all
possible states will be sampled with equal probability.

\section{Monte Carlo algorithms for the three colour model}
\label{mctc}
We now have two Monte Carlo algorithms which correctly sample the states of
the square ice model and since, as we showed in
Section~\sref{threecolouring}, the states of this model can be mapped onto
the states of the three colour lattice model, we can of course use the same
algorithm to study the three colour model.  In this section however, we
will explore the other side of the same question: is there a natural Monte
Carlo dynamics for the three-colouring model which could then be used to
sample the states of the ice model?  It turns out that there is, and the
resulting algorithm provides not only an efficient way of simulating ice
models, but will also prove useful when we get onto the energetic ice
models of Section~\sref{energies} in which different types of vertices are
assigned different energies.

In the three-colouring representation the degrees of freedom---the
colours---are located on the plaquets of the lattice, rather than at the
vertices, and, as we showed earlier, the ice rules translate into the
demand that nearest-neighbour squares have different colours.  Just as in
the case of the square ice model, there is no obvious update move which
will take us from state to state.  Although there are some states in which
the colour of one square can be changed from one value to another without
violating the ice rules, there are also states in which no such moves are
possible, and therefore single-plaquet moves of this kind cannot reach
these states, and so do not lead to an ergodic dynamics.  Again then, we
must resort to non-local moves, and the most obvious such move is to look
for clusters of nearest-neighbour plaquets of only two colours, call them
$A$ and $B$, entirely surrounded by plaquets of the third colour $C$.  A
move which exchanges the two colours $A$ and $B$ in such a cluster but
leaves the rest of the lattice untouched satisfies the ice rules, and this
suggests the following cluster-type algorithm for square ice:
\begin{enumerate}
\item We choose a plaquet at random from the lattice as the seed square for
  the cluster.  Suppose this plaquet has colour $A$.
\item We choose another colour $B\ne A$ at random from the two other
  possibilities.
\item Starting from our seed square, we form a cluster by adding all
  nearest-neighbour squares which have either colour $A$ or colour $B$.  We
  keep doing this until no more such nearest neighbours exist.
\item The colours $A$ and $B$ of all sites in the cluster are exchanged.
\end{enumerate}

There are a couple of points to notice about this algorithm.  First, the
cluster possesses no nearest neighbours of either colour $A$ or colour $B$
and therefore all its nearest neighbours must be of the third colour, $C$.
In the simplest case, the seed square has no neighbours of colour $B$ at
all, in which case the cluster consists of only the one plaquet.  It is
crucial to the working of the algorithm that such moves should be possible.
If we had chosen instead to seed our cluster by picking two neighbouring
plaquets and forming a cluster with their colours, single-plaquet moves
would not be possible and we would find that the algorithm satisfied
neither ergodicity nor detailed balance.  Notice also that within the
boundary of colour $C$, the cluster of $A$s and $B$s must form a
checkerboard pattern, since no two $A$s or $B$s can be neighbours.

We are now in a position to prove that our algorithm satisfies the
conditions of ergodicity and detailed balance.  In this case it turns out
that detailed balance is the easier to prove.  Consider, as before, a Monte
Carlo move which takes us from a state $\mu$ to a state $\nu$, and suppose
that this move involves a cluster of $m$ squares.  The probability of
choosing our seed square in this cluster is $m/N$, where $N$ is the total
number of plaquets on the lattice.  The probability that we then choose $B$
as the other colour for the cluster is $\frac12$, and after that there are
no more choices: the algorithm specifies exactly how the cluster should be
grown from here on.  Thus the total probability for the move from $\mu$ to
$\nu$ is $m/(2N)$.  Exactly the same argument applies for the reverse move
from $\nu$ to $\mu$ with the same values of $m$ and $N$, and hence the
rates for forward and reverse moves are the same.  Thus detailed balance is
obeyed.

The proof of ergodicity is a little trickier.  It involves two steps.
First, we show that from any configuration we can evolve via a finite
sequence of reversible moves to a checkerboard colouring (a configuration
in which one of the three colours is absent).  Then we show that all
checkerboard colourings are connected through reversible moves.

Any configuration of the lattice can be regarded as a number of
checkerboard regions consisting of only two colours, divided by boundaries.
This result is obvious, since each site of colour $A$ must have at least
two neighbours with the same colour, and therefore each square on the
lattice belongs to a checkerboard domain of at least three squares.
However, under the dynamics of our proposed Monte Carlo algorithm, the
boundaries between these domains can move.  If we have a domain of colours
$A$ and $B$ and another of $B$ and $C$ then by choosing one of the plaquets
on the boundary as the seed square for our Monte Carlo move, and $B$ as one
of the colours for the cluster, we can make the boundary move one square in
one direction or the other, with the direction depending on whether the
other colour for the cluster was $A$ or $C$.  In this way we can take a
single simply-connected cluster of one checkerboard pattern and, over a
number of steps, grow its border until the cluster covers the entire
lattice, leaving the lattice in a checkerboard state.

There are six of these checkerboard colourings, and from any one of them
the others can easily be reached, since on a checkerboard the colour of any
square can be changed on its own without changing any other squares.  Thus
for example we can get from a checkerboard of colours $A$ and $B$ to one of
$A$ and $C$ by changing all the $B$s to $C$s one by one.  All other
combinations can be reached by a similar process.

Since we can get from any state $\mu$ to a checkerboard colouring and from
any checkerboard to any other, all via reversible moves, it follows that our
algorithm is ergodic.

The algorithm presented above, a single-cluster algorithm, resembles in
spirit the Wolff single-cluster algorithm for the Ising
model~\cite{Wolff89}.  It is also possible to construct a multi-cluster
algorithm for the three-colouring model, similar to the Swendsen-Wang
algorithm for the Ising model~\cite{SW87}. In this algorithm we start by
choosing at random a pair of colours $A$ and $B$.  Then we construct all
clusters of nearest-neighbour spins made out of these two colours, and for
each cluster we choose at random with 50\% probability whether to exchange
the two colours or not.  This algorithm satisfies ergodicity for the same
reason the single-cluster algorithm did---we can repeatedly choose two
colours for the move until a single cluster grows to fill the entire
lattice, giving a checkerboard pattern.  But we can get from any
checkerboard to any other so that any state can be reached in a finite
number of steps on a finite lattice.  The algorithm also satisfies detailed
balance: the probability of selecting a particular two out of the three
colours for a move is $\frac13$, and the probability of exchanging the
colours in a particular set of clusters is $2^{-n}$, where $n$ is the
number of clusters.  The probability for the reverse move is exactly the
same, and hence detailed balance is upheld.

\section{Comparison of algorithms for square ice}
\label{squaresim}
In the previous sections, we have proposed four algorithms for the
simulation of square ice: the long loop algorithm, the short loop
algorithm, the single-cluster three-colouring algorithm, and the
full-lattice three-colouring algorithm.  In this section we consider these
algorithms one by one and compare their computational efficiency.

\begin{figure}
\normalfigure{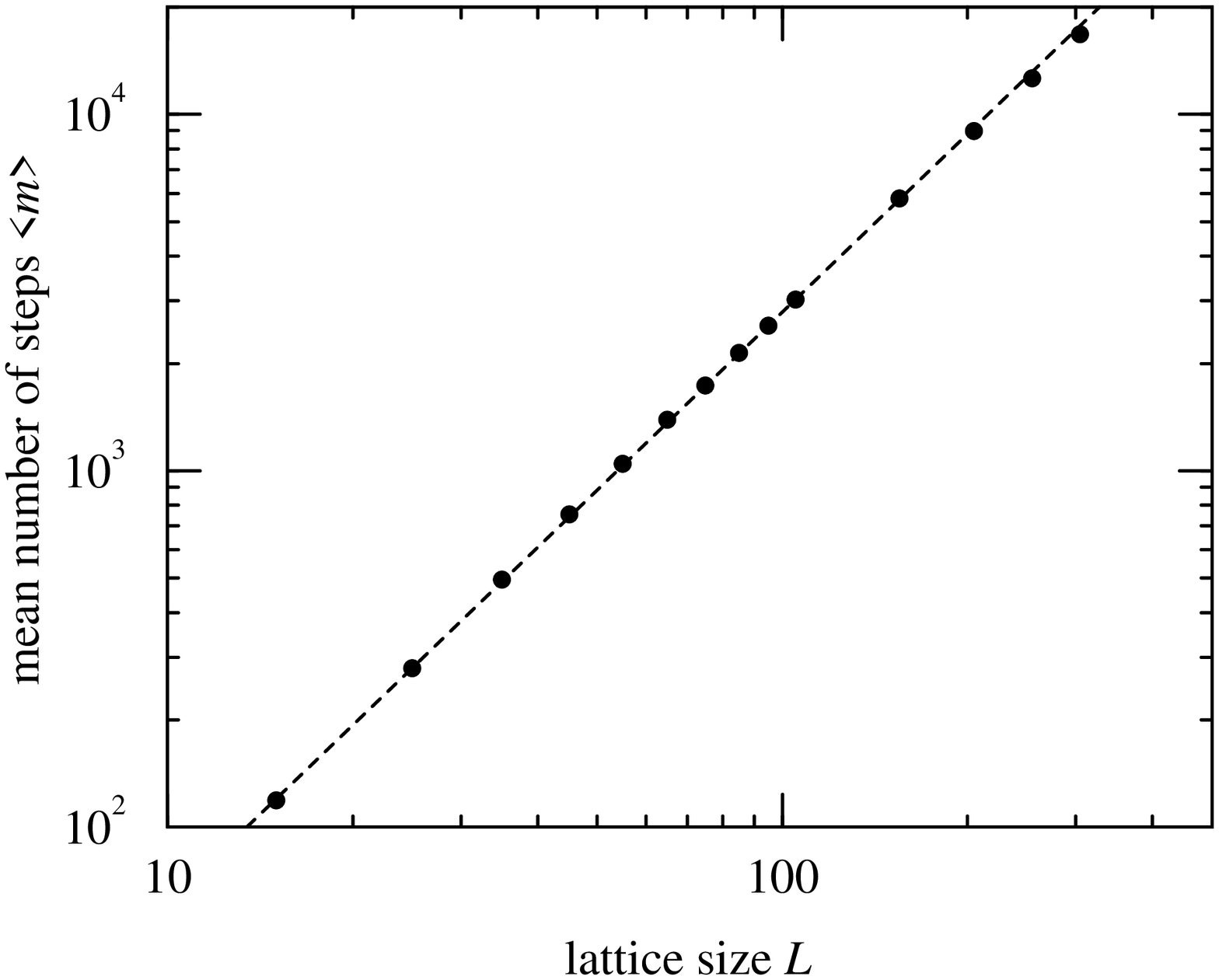}
\capt{The mean length $\langle m \rangle$ of loops in the long loop
  algorithm as a function of system size $L$.  We find that $\langle m
  \rangle \sim L^{1.665\pm0.002}$.}
\label{walk}
\end{figure}

\begin{figure}
\normalfigure{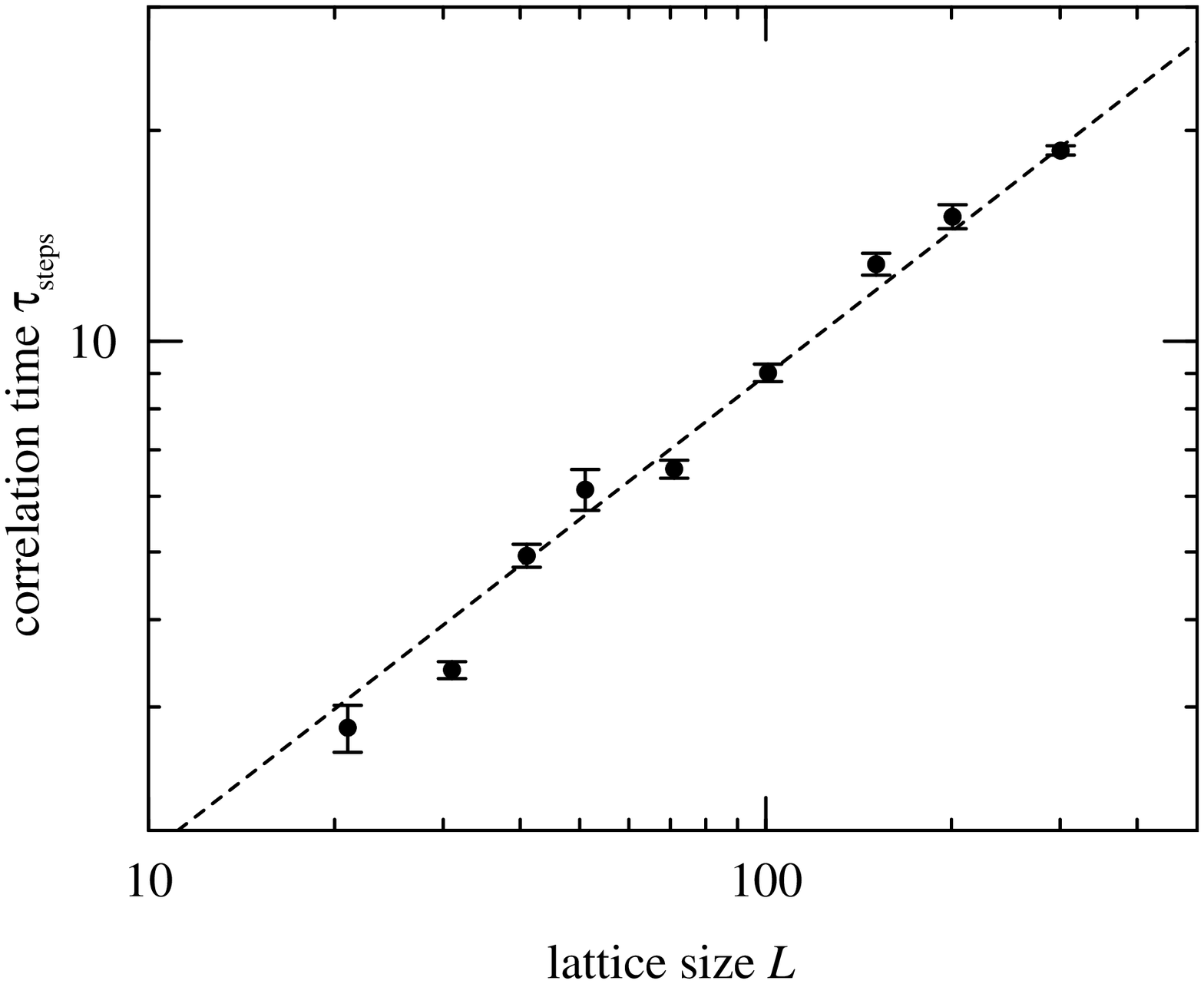}
\capt{The correlation time in Monte Carlo steps of the long loop
  algorithm as a function of system size $L$.  The best fit
  straight line gives $\tau_{\rm steps} \sim L^{0.68\pm0.03}$.}
\label{longloop}
\end{figure}

\subsection{The long loop algorithm}
The long loop algorithm involves the creation of a pair of ionic defects,
one of which diffuses around the lattice until it recombines with the
first, in the process reversing all the arrows along the path of its
diffusion.  To assess the efficiency of this algorithm, we first measure
the average number of steps which the wandering defect takes before it
recombines as a function of the system size $L$.  For an ordinary random
walker on a square lattice, this number scales as $L^2$.  In the case of
the wandering defect however, we find that it scales instead as
$L^{1.67}$---see Figure~\fref{walk}.  The amount of CPU time required per
step in our algorithm increases linearly with the size of the loop, and
hence we expect the CPU time per Monte Carlo step also to increase with
system size as $L^{1.67}$.  This is not necessarily a problem; since longer
loops reverse more arrows as well as taking more CPU time it is unclear
whether longer is better in this case, or worse.  To answer this question
we need to consider the correlation time of the algorithm.  We have
measured the correlation time for an observable $\rho_{\rm sym}$ which we
define to be the density of the symmetric vertices~5 and~6 in
Figure~\fref{sixvertex}.  As Figure~\fref{longloop} shows, when we measure
time in Monte Carlo steps we find a correlation time $\tau_{\rm steps} \sim
L^{0.68\pm0.03}$.  It is however more common (and more convenient for the
comparison of our algorithms) to measure time in ``sweeps'' of the lattice,
which in this case means arrow flips per bond on the lattice.  On average,
each Monte Carlo step corresponds to $\langle m \rangle/(dL^d)$ sweeps on a
$d$-dimensional lattice, which means that the correlation time on our 2D
lattice goes as
\begin{equation}
\tau \sim L^{0.68} {L^{1.67}\over L^2} = L^{0.35\pm0.03}.
\end{equation}
This quantity measures the amount of computer effort we have to invest, per
unit area of the lattice, in order to generate an independent configuration
of the arrows. 

The square ice model is a critical model, possessing an infinite
correlation length~\cite{Baxter82}.  Thus it comes as no surprise that the
correlation time scales as a non-integral power law with system size.  The
exponent $z=0.35$ is the dynamic exponent for the critical system---the
anomalous scaling of the correlation time over and above the $L^d$ scaling
expected of a system far from criticality.  As dynamic exponents go, this
is a reasonably small one.  The Metropolis algorithm for the normal Ising
model in two dimensions for example has a dynamic exponent of about
$z=2.17$~\cite{NB96}, making simulations of the model very time consuming
for large lattices close to criticality.  However, as we will see, some of
our other algorithms for square ice do better still, possessing dynamic
exponents not measurably different from zero.

\subsection{The short loop algorithm}
\label{shortsec}
The short loop algorithm of Section~\sref{smallloops} also involves
creating a pair of defects and having one of them diffuse around.  Recall
however, that in this case the wandering defect only has to find {\em
  any\/} of the sites which it has previously visited in order to close the
loop and finish the Monte Carlo step.  If the diffusion were a normal
random walk then this process would generate loops of a finite average
length.  Although the diffusion of defects in square ice is not a true
random walk it turns out once more that the same result applies.
Numerically we find that the average number of steps per move is $\langle m
\rangle=13.1$, independent of the lattice size, for a sufficiently large
lattice.  This figure includes the steps taken at the end of the move which
simply flip a number of arrows back to their starting configuration and
therefore have no net effect on the state of the system.  (See
Section~\sref{smallloops}.)  We find that typically about 58\% of the
arrows reversed during a move have to be restored in their original state.
This is certainly a source of inefficiency in the algorithm.

\begin{figure}
\normalfigure{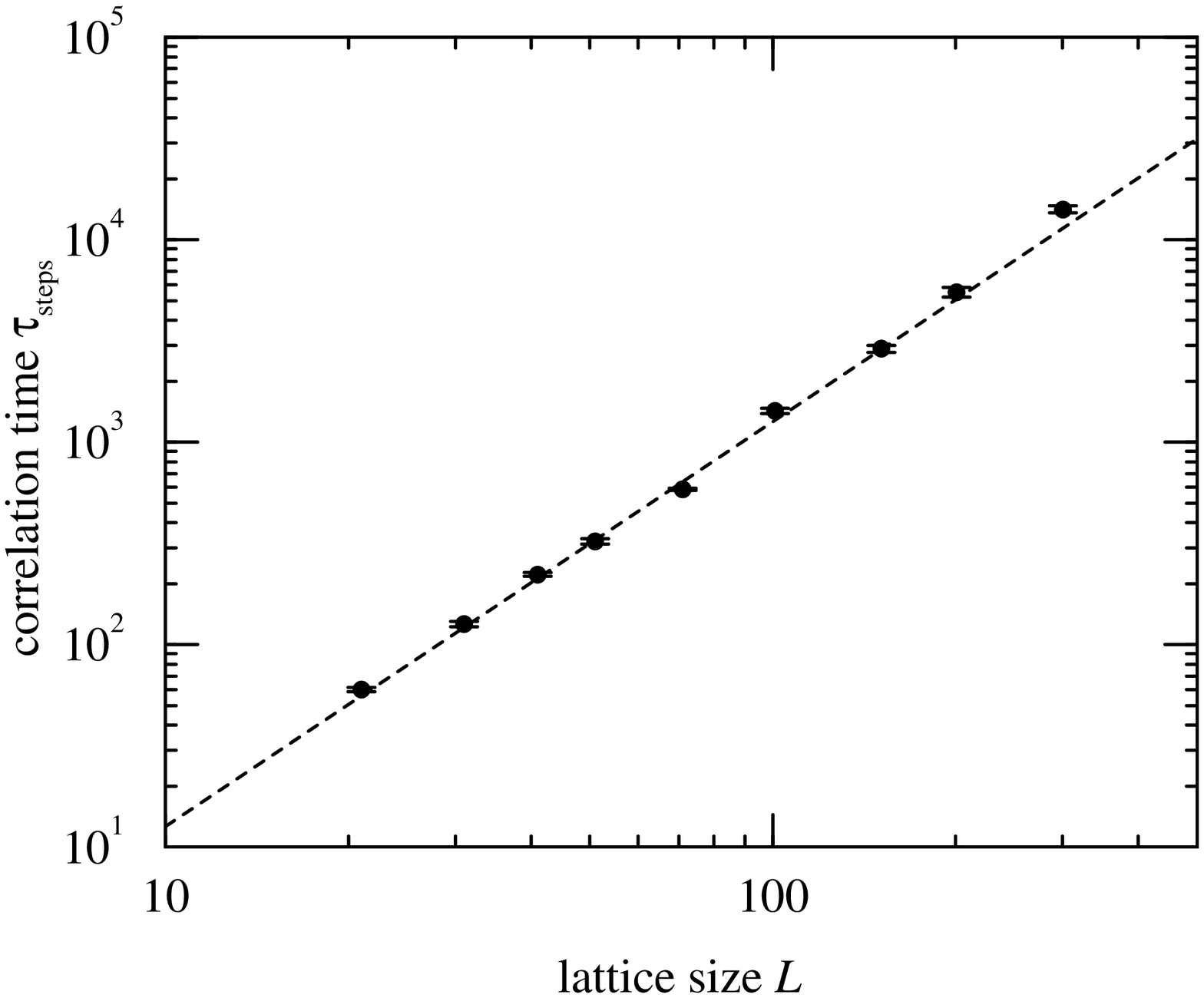}
\capt{The correlation time $\tau_{\rm steps}$ of the short loop algorithm
  measured in Monte Carlo steps as a function of system size.  The best fit
  straight line gives $\tau_{\rm steps} \sim L^{2.00\pm0.01}$.}
\label{shortloop}
\end{figure}

The correlation time measured in Monte Carlo steps $\tau_{\rm steps}$, for
the same observable $\rho_{\rm sym}$ as above, increases as $L^2$
(Figure~\fref{shortloop}).  Since the mean number of steps in a loop is
independent of $L$, the correlation time per unit volume goes as
\begin{equation}
\tau \sim L^2 {L^0\over L^2} = \mbox{constant}.
\end{equation}
Thus the short loop algorithm scales optimally with system size.  To the
accuracy of our simulations the dynamic exponent is $z=0.00\pm0.01$

\subsection{Single-cluster three colouring algorithm}
Our third algorithm is the single-cluster three-colouring algorithm
outlined in Section~\sref{threecolouring}.  For this algorithm the average
CPU time per Monte Carlo step scales as the average cluster size $\langle c
\rangle$.  Like the loop length in the long loop algorithm, this quantity
scales up with increasing lattice size and numerically we find that
\begin{equation}
\langle c \rangle \sim L^{1.5}.
\end{equation}
The correlation time per Monte Carlo step goes as
\begin{equation}
\tau_{\rm steps} \sim L^{1.8},
\end{equation}
and hence the correlation time in steps per site goes as
\begin{equation}
\tau \sim L^{1.8} {L^{1.5}\over L^2} = L^{1.3},
\end{equation}
indicating that the single-cluster algorithm is a very poor algorithm
indeed for studying square ice on large lattices.

\begin{figure}
\normalfigure{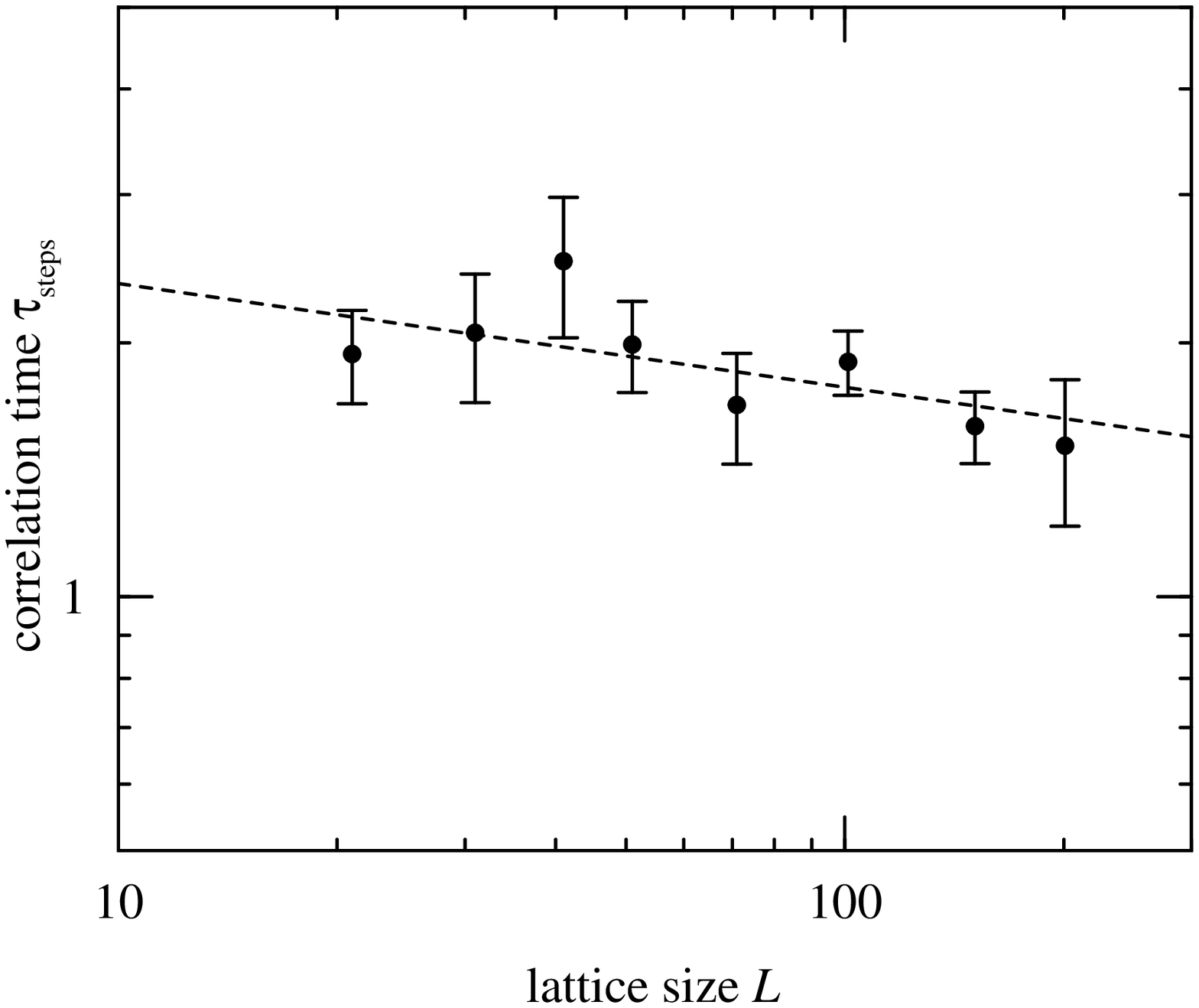}
\capt{The correlation time $\tau_{\rm steps}$ of the full-lattice
  three-colouring algorithm measured in Monte Carlo steps as a function of
  system size.  The best fit straight line gives $\tau_{\rm steps} \sim
  L^{-0.12\pm0.07}$.}
\label{sw3sqr}
\end{figure}

\subsection{Full-lattice three colouring algorithm}
\label{fltcsq}
Our last algorithm, the full-lattice three colouring algorithm, also
described in Section~\sref{threecolouring}, generates clusters in a way
similar to the single cluster algorithm, but rather than generating only
one cluster per Monte Carlo step, it covers the whole lattice with them.
For this algorithm we find numerically that the correlation time $\tau_{\rm
  steps}$ measured in Monte Carlo steps is approximately constant as a
function of lattice size (Figure~\fref{sw3sqr}).  Since each Monte Carlo
move updates sites over the entire lattice, the CPU time per move scales as
$L^2$ and hence the correlation time in moves per site is
\begin{equation}
\tau \sim L^{0} {L^2\over L^2} = L^0.
\end{equation}
Thus, like the short loop algorithm, this one possesses optimal scaling as
lattice size increases, with a measured dynamic exponent of
$z=-0.12\pm0.07$.

Comparing the four algorithms, clearly the most efficient ones for large
systems are the short loop algorithm and the full-lattice three-colouring
algorithm.  In both other algorithms, the computer time required to
generate an independent configuration of the lattice increases with system
size.  The larger impact of the larger moves in these algorithms does not
compensate for the extra effort invested generating them.  Between the
short loop algorithm and the full-lattice three-colouring algorithm, it is
harder to decide the winner, since both have the same scaling of CPU
requirements with system size.  Our results show in fact that the two
algorithms are comparable in speed, both giving on the order of a million
site updates per second on the workstations used for this study.  The loop
algorithm is perhaps slightly faster (maybe 10 or 20 per cent) and has the
advantage of working on lattices of other topologies as well as the square
lattices used here.  The three-colouring algorithm on the other hand is
considerably more straightforward to program.

\begin{figure}
\normalfigure{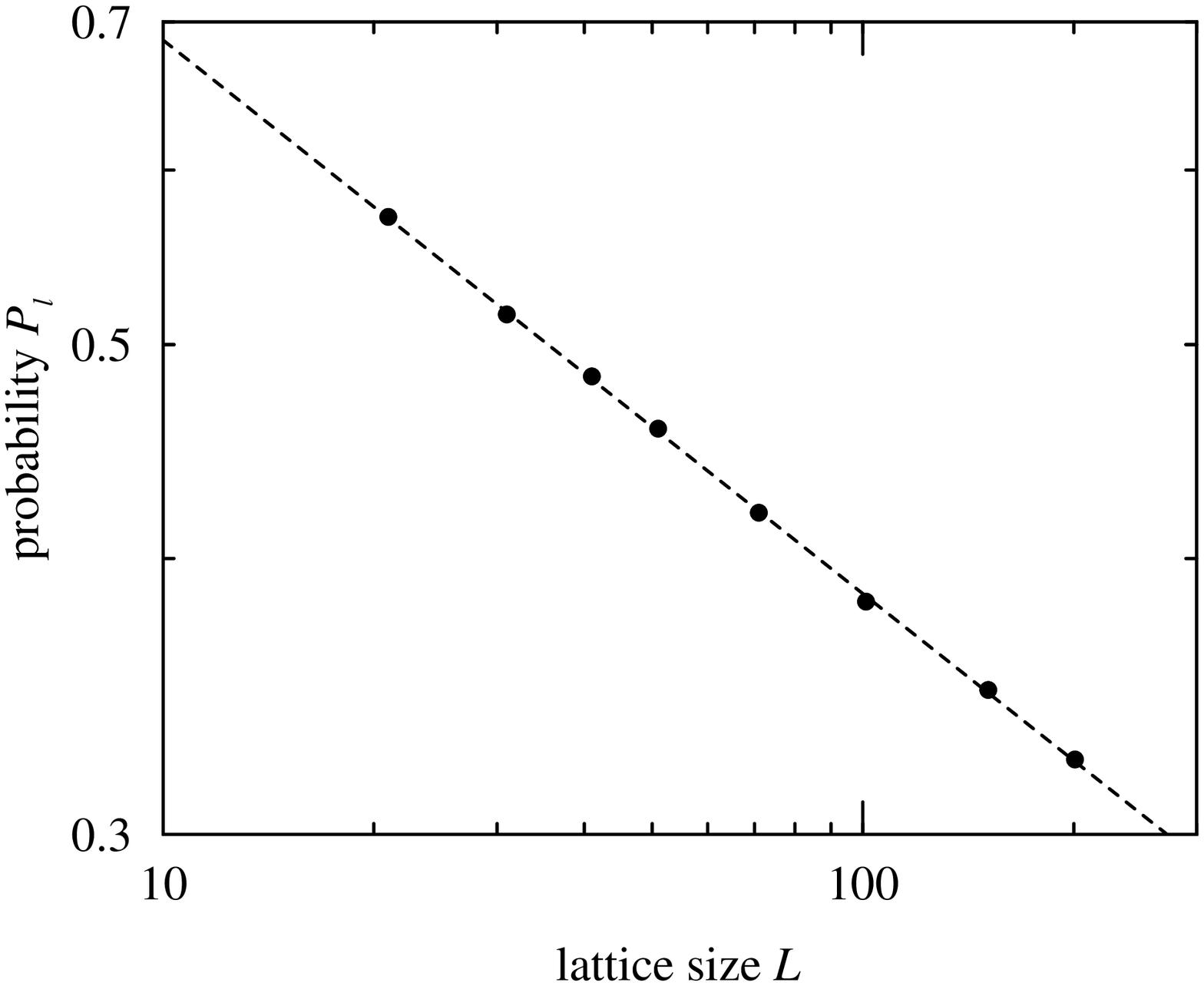}
\capt{The probability $P_l$ that a site belongs to the longest loop in the
  fully-loop-covered representation of square ice, as a function of system
  size $L$.  We find that $P_l \sim L^{-0.251\pm0.002}$.}
\label{squareloop}
\end{figure}

As an example of the use of our algorithms, we have measured one of the
simplest non-trivial critical exponents for the square ice model.  As we
showed in Section~\sref{fullloop}, each state of the square ice model
corresponds to a configuration of a square lattice which is entirely
covered by closed, non-self-intersecting loops.  Using our full-lattice
three-colouring algorithm, we have measured the probability $P_l$ that a
particular site is visited by the largest loop in such a model as a
function of lattice size $L$.  The results are shown in
Figure~\fref{squareloop}.  The data closely follow a power-law: $P_l \sim
L^{-0.25}$.

\section{Energetic ice models}
\label{energies}
There are a number of other systems besides H$_2$O with four-fold
coordinated hydrogen bonds, the most studied being potassium dihydrogen
phosphate (KH$_2$PO$_4$), also known as KDP.  Slater~\cite{Slater41} argued
that KDP at low temperatures could be modeled using a six-vertex model in
which vertices~1 and~2 in Figure~\fref{sixvertex} are favoured by giving
them an energy $-\epsilon$, while all the others are given energy zero.
Notice that it is possible to form a domain on a square lattice consisting
only of type~1 vertices, or only of type~2.  Thus there are two degenerate
ground states of the KDP model in which the lattice is entirely covered
with vertices of one of these two types, and the model displays a
symmetry-breaking phase transition from a high-temperature phase in which
the two appear with equal probability to a low-temperature one in which one
or the other dominates.  A suitable order parameter to describe this
transition is the polarization, or average direction of the arrows:
\begin{equation}
{\bf P} = {1\over\sqrt{2}N} \sum_i \hat{{\bf n}}_i,
\label{polarization}
\end{equation}
where the vector $\hat{{\bf n}}_i$ is a unit vector in the direction of the
\th{i} arrow.  In the thermodynamic limit the polarization will be zero
above the critical temperature $T_c$, and non-zero below it with a
direction either upwards and to the right, or downwards and to the left,
and a magnitude which approaches unity as $T\to0$.

Another widely-studied energetic ice model is the so-called F model
\cite{Rys63}, in which vertices~5 and~6 in Figure~\fref{sixvertex} are
given a lower energy $-\epsilon$ and all the others are given energy zero.
This model has a ground state in which vertices~5 and~6 alternate in a
checkerboard pattern across the lattice.  There are again two possible such
ground states, depending on which type of vertex falls on the even sites of
the checkerboard and which on the odd, and there is a symmetry breaking
phase transition from the high-temperature phase in which the two vertices
fall on even and odd sites with equal probability.  Since neither vertex~5
nor vertex~6 possesses any net polarization, the value of ${\bf P}$ is zero
in the thermodynamic limit for the F model, regardless of temperature.
However, one can define an anti-ferroelectric order parameter which does
become non-zero in the low-temperature phase~\cite{Lieb67a,Lieb67b}.

A third energetic ice model which has attracted some attention recently is
the staggered, body-centred solid-on-solid~(BCSOS)
model~\cite{Knops79,CMB97}.  In this model the square lattice is divided
into even and odd sites and the vertex types are divided into three groups.
On even lattice sites vertices of types~1 and~2 have energy $\epsilon$ and
type~3 and~4 have energy $\epsilon'$, on odd lattice sites $\epsilon$ and
$\epsilon'$ are reversed, and vertices of types~5 and~6 have energy zero
everywhere.  The values of $\epsilon$ and $\epsilon'$ may be either
positive or negative.  In the height representation described in
Section~\sref{randomsurfaces} this model is believed to described
roughening transitions in certain ionic crystals with the CsCl structure.

\subsection{Monte Carlo algorithms for energetic ice models}
In Section~\sref{squaresim} we developed a variety of elementary ergodic
moves for sampling the states of ice models on square lattices, and showed
how these could be used to create Monte Carlo algorithms for the square ice
model, in which all states have the same energy.  We can use the same sets
of elementary moves to create Monte Carlo algorithms for the energetic ice
models as well.  The simplest method is to employ a Metropolis-type scheme
in which instead of always carrying out every move generated by the
algorithm, we carry them out with an acceptance probability $P$ which
depends on the energy difference $\Delta E = E_\nu - E_\mu$ between the
states $\mu$ and $\nu$ of the system before and after the move:
\begin{equation}
P = \Biggl\lbrace \begin{array}{ll}
    \e^{-\beta\Delta E}\qquad & \mbox{if $\Delta E>0$}\\
    1 & \mbox{otherwise.}
    \end{array}
\end{equation}
Here we give examples of algorithms for the F model, but the same ideas can
easily be adapted for use with other energetic ice models.

The Hamiltonian of the F model is given by
\begin{equation}
\label{Fhamiltonian}
H = - \epsilon \sum_i \left( \delta_{v_i,5} + \delta_{v_i,6} \right),
\end{equation}
where $v_i$ is a number corresponding to the type of vertex at site $i$,
using the numbering scheme illustrated in Figure~\fref{sixvertex}.

\begin{figure}
\sidefigure{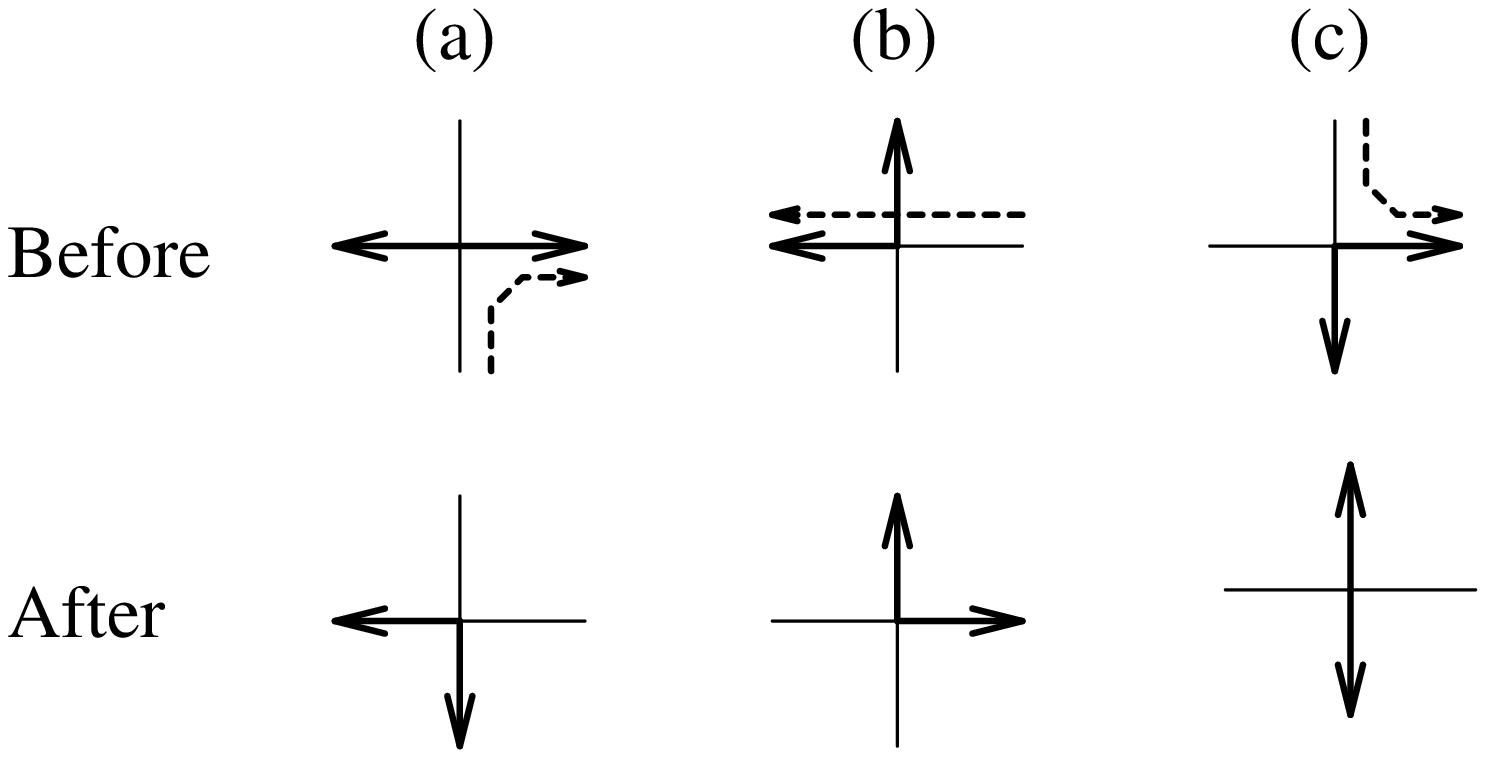}{Symmetric vertices become non-symmetric if a loop
  passes through them (a). Non-symmetric vertices stay non-symmetric if the
  loop through them goes straight through (b), but become symmetric if the
  loop makes a turn (c).}
\label{loopdir}
\end{figure}

Let us first consider algorithms in which the proposed moves involve
reversing the directions of the arrows around a loop on the lattice, as in
the long and short loop algorithms of Sections~\sref{longloops}
and~\sref{smallloops}.  For these moves the only vertices which change type
(and hence energy) are those which the loop passes through.  As is shown in
Figure~\fref{loopdir}, a symmetric vertex (type~5 or~6) always becomes
non-symmetric if the loop passes through it, thereby increasing the total
energy.  If the loop passes straight through a non-symmetric vertex, the
vertex remains non-symmetric and its energy is unchanged.  On the other
hand, if the loop makes a turn as it passes through a non-symmetric vertex,
the vertex becomes symmetric and the energy decreases.  Thus, given a
particular loop, we can calculate the value of $\Delta E$ by counting the
number $m$ of symmetric vertices which the loop passes through and the
number $n$ of non-symmetric vertices in which it makes a $90^\circ$ turn,
and applying the formula
\begin{equation}
\Delta E=(m-n) \epsilon.
\end{equation}

\begin{figure}
\normalfigure{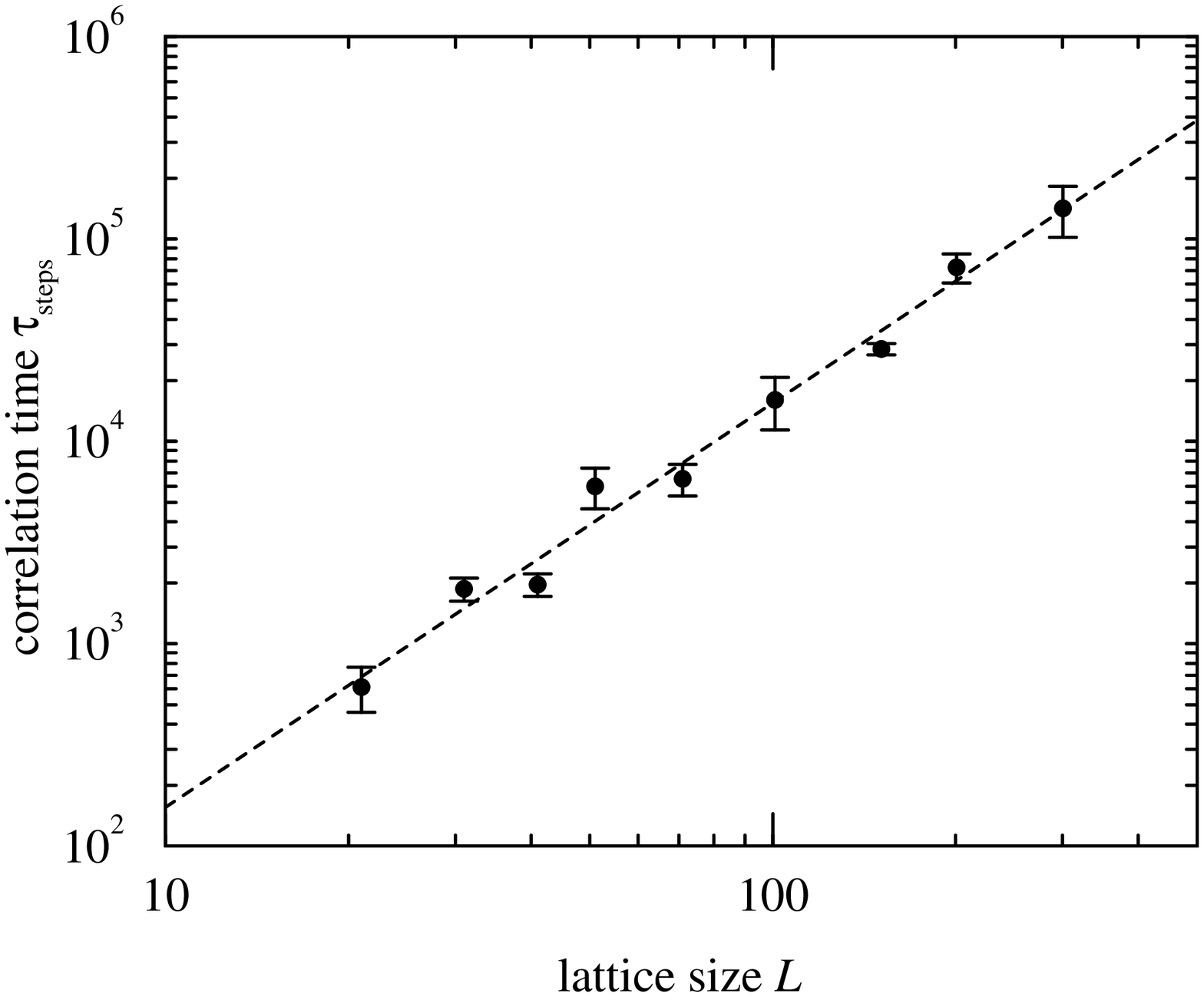}
\capt{The correlation time $\tau_{\rm steps}$ of the short loop algorithm
  for the F model measured in Monte Carlo steps, as a function of system
  size.  The best fit straight line gives $\tau_{\rm steps} \sim
  L^{2.00\pm0.09}$.}
\label{slcrit}
\end{figure}

The density of symmetric vertices in the F model increases with decreasing
temperature, so that the average number of symmetric vertices through which
a loop passes grows as we go to lower temperatures.  Since each symmetric
vertex which we pass adds an amount $\epsilon$ to $\Delta E$, it is clear
that loop moves will carry an energy cost which increases with their length
and that long loops will be very energetically costly, especially at low
temperatures.  This suggests that the short loop algorithm of
Section~\sref{smallloops} will be more efficient for the simulation of the
F model at finite temperature.  In Figure~\fref{slcrit} we show the
correlation time $\tau_{\rm steps}$ measured in Monte Carlo steps for this
algorithm, and the best fit to these data gives us
\begin{equation}
\tau_{\rm steps} \sim L^{2.0}.
\end{equation}
As with square ice, the number of sites updated by a single Monte Carlo
step tends to a constant for large lattices, so that the correlation time
in steps per site is
\begin{equation}
\tau \sim L^{2.0} {L^0\over L^2} = L^0.
\end{equation}
To the accuracy of our simulations then, this algorithm has a zero dynamic
exponent.  However, it turns out that this algorithm is still quite
inefficient for temperatures in the region of the critical temperature and
below.  For example at $T_c$ the acceptance ratio is 36\% so that nearly
two thirds of the computational effort is wasted.  For this reason we have
investigated a number of other algorithms for simulating the F model.

How can we increase the acceptance ratio of our Monte Carlo algorithm?  We
would like to propose moves that are less likely to cost energy.  For
example, if we can encourage the loop to make turns in non-symmetric
vertices, we will on average end up with a lower final energy, since a
reversal of the arrows around the loop will create more symmetric vertices.
Unfortunately, it turns out to be quite complicated to formulate a correct
algorithm along these lines, and the expression for the acceptance ratio
becomes quite tedious.  There is however an elegant alternative, which is
to employ a three-colouring algorithm of the type discussed in
Section~\sref{mctc}.

The equivalent of a symmetric vertex in the three-colouring model is a
group of four squares in which both of the diagonally opposing pairs share
the same colour.  In non-symmetric vertices only one of these two diagonal
pairs share the same colour.  Making use of this observation we can write
the Hamiltonian of the F-model (Equation~\eref{Fhamiltonian}) in the form
\begin{equation}
\label{3colourE2}
H = - \epsilon \sum_{[i,j]} (\delta_{c_i,c_j} - \frac12)
  = N\epsilon - \epsilon \sum_{[i,j]} \delta_{c_i,c_j},
\end{equation}
where the summation runs over all pairs of next-nearest-neighbour squares
$[i,j]$, and $c_i$ is the colour of square $i$.  We see that it is
energetically favourable to have pairs of next-nearest-neighbour squares
with an identical colour.  We can make use of this observation to create an
efficient algorithm for the three-colouring model.  In this algorithm, as
in the algorithms for square ice discussed in Section~\sref{mctc}, we
build clusters of nearest-neighbour plaquets of two colours, but now in
addition, we also add to the cluster next-nearest-neighbour plaquets as
well.  In detail our algorithm is as follows:
\begin{enumerate}
\item We choose a plaquet at random from the lattice as the seed square for
  the cluster.  Suppose that this plaquet has colour $A$.
\item We choose another colour $B\ne A$ at random from the two other
  possibilities.
\item Starting from our seed square, we form a cluster by adding all
  nearest-neighbour squares which have either colour $A$ or colour
  $B$, and in addition we now also add to the cluster the squares
  which are next-nearest neighbours of some square $i$ which is
  already in the cluster, provided they have the {\em same\/} colour
  as square $i$.  However, we make this latter addition with a
  temperature-dependent probability $\alpha<1$, whose value we
  calculate below in order to satisfy the condition of detailed
  balance.  We go on adding squares to the cluster in this way until
  no more additions are possible.
\item The colours $A$ and $B$ of all sites in the cluster are exchanged.
\end{enumerate}

We can also make a full-lattice version of this algorithm in exactly the
same way as for the square ice case.  We choose two colours $A$ and $B$ at
random and create clusters all over the lattice from these two, using the
method above.

It is straightforward to prove ergodicity for these algorithms.  Since our
three-colouring algorithms for square ice were ergodic (see
Section~\sref{mctc}), and since each move in the square ice algorithms is
also a possible move in our F model algorithm (as long as $\alpha<1$), the
result follows immediately.

Detailed balance is a little more tricky.  We outline the argument here for
the single-cluster version of the algorithm.  As before, consider two
states $\mu$ and $\nu$ which differ by the exchange of colours in a single
cluster of $m$ squares.  The probability of choosing the seed square in
this cluster is $m/N$ and the probability that we choose the correct second
colour to create this particular cluster is $\frac12$, just as in the
square ice case.  However, we now also have a factor of $\alpha$ for every
square which we add to the cluster which is only a next-nearest neighbour
of another and not a nearest neighbour.  And we have a factor of $1-\alpha$
for every such site which we could have added but didn't.  Thus the overall
probability of making the move from $\mu$ to $\nu$ is
\begin{equation}
P(\mu \rightarrow \nu) = {m\over2N} \prod_{[i,j]_{\rm con}} \alpha
\prod_{[i,j]_{\rm dis}} (1-\alpha)^{\delta(c_i^{(\mu)},c_j^{(\mu)})},
\end{equation}
where the two products run over pairs of next-nearest neighbours which are
connected to or disconnected from the cluster respectively.  We will find
it easier to work with the logarithm of this probability:
\begin{eqnarray}
\log P(\mu \rightarrow \nu) = -\log(m/2N) &+&
\log\alpha \sum_{[i,j]_{\rm con}} 1\nonumber\\
&+& \log(1-\alpha) \sum_{[i,j]_{\rm dis}} \delta(c_i^{(\mu)},c_j^{(\mu)}).
\end{eqnarray}
The expression for $\log P(\nu \rightarrow \mu)$ is identical except for
the exchange of the labels $\mu$ and $\nu$.

We want to know the ratio of the probabilities for the forward and reverse
moves:
\begin{equation}
\label{logtransratio}
\log \frac{P(\mu \rightarrow \nu)}{P(\nu \rightarrow \mu)} =
\log(1-\alpha) \sum_{[i,j]_{\rm dis}}
\delta(c_i^{(\mu)},c_j^{(\mu)}) - \delta(c_i^{(\nu)},c_j^{(\nu)}).
\end{equation}
The energy difference $\Delta E$ between states $\mu$ and $\nu$ is equal to
$\epsilon$ times the change in the number of identically coloured
next-nearest-neighbour squares (see Equation~\eref{3colourE2}).  The only
contribution to this sum comes from next-nearest-neighbour pairs $[i,j]$
such that $i$ belongs to the cluster and $j$ does not, since all other
pairs contribute the same amount to the Hamiltonian in state $\mu$ as in
state $\nu$.  Thus
\begin{equation}
\label{logboltz}
\Delta E = E_\nu - E_\mu = - \epsilon \sum_{[i,j]_{\rm dis}}
[ \delta(c_i^{(\nu)},c_j^{(\nu)})-\delta(c_i^{(\mu)},c_j^{(\mu)}) ].
\end{equation}
In order to satisfy the condition of detailed balance we want the ratio of
the rates $P(\mu\to\nu)$ and $P(\nu\to\mu)$ to be equal to the ratio
$\exp(-\beta\Delta E)$ of the Boltzmann weights of the two states.
Comparing Equations~\eref{logtransratio} and~\eref{logboltz}, we see that
this can be arranged by setting $\log(1-\alpha) = -\beta\epsilon$, or
\begin{equation}
\alpha = 1 - \e^{-\beta\epsilon}.
\end{equation}

The proof of detailed balance for the full-lattice version of the
algorithm follows from the single-cluster version just as in the case of
the square ice model.

\begin{figure}
\normalfigure{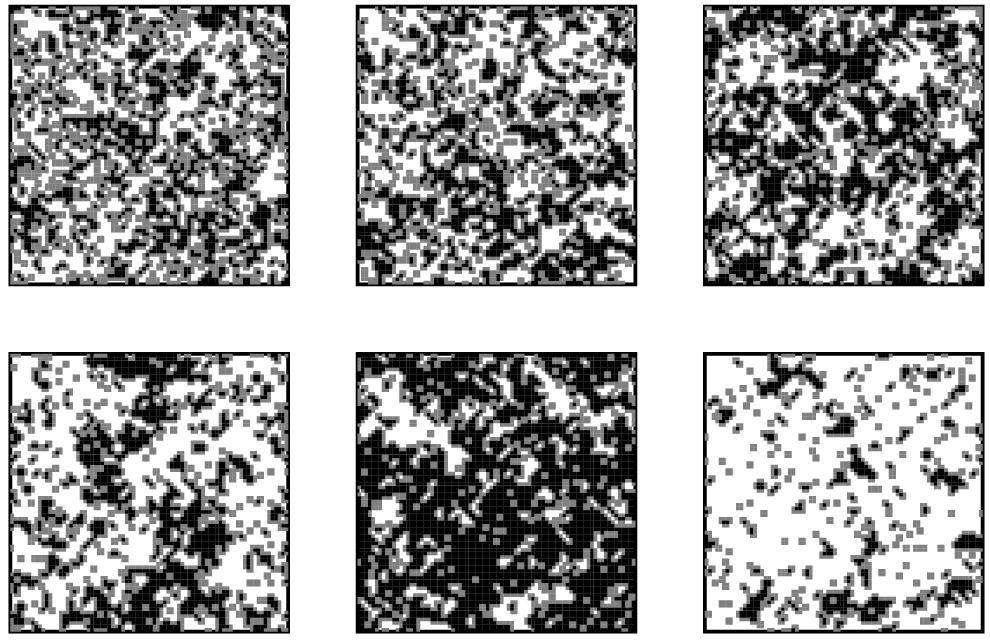}
\capt{Sample configurations of the F-model for increasing $\beta$.  Grey
  squares denote vertices of types 1, 2, 3, and 4. White vertices denote
  either vertices of type 5 on even lattice sites, or vertices of type 6 on
  odd lattice sites. Other vertices are black.  Top row: $\beta/\beta_c=$
  0.5, 0.8, and 0.9.  Bottom row: $\beta/\beta_c=$ 1.0, 1.1, and 1.2.}
\label{Tdep}
\end{figure}

In Figure~\fref{Tdep} we show some results of simulations of the F model
using the full-lattice version of the algorithm described above.  In this
figure we have coloured areas of the two low energy domains (checkerboards
of symmetric vertices) in black and white---type~5 vertices on even lattice
sites and type~6 vertices on odd lattice sites are black, while type~6
vertices on even lattice sites and type~5 vertices on odd lattice sites are
white.  All other vertices are in grey.

The phase transition is clearly visible in the figure as a change from a
state in which black and white appear with equal frequency to one in which
one or the other dominates.  Analytically it is known that this transition
takes place at $T_c=\epsilon/\ln 2$.  This number is rather difficult to
measure numerically however, since the phase transition is of infinite
order; no matter how often you differentiate the energy or the density of
symmetric vertices with respect to temperature, you will not see a
singularity.  Nonetheless there is a phase transition.  For instance, the
absolute value of the difference in density of black and white squares on
an infinite lattice is strictly zero above the critical temperature, while
non-zero below, ruling out any analytic behaviour.

\begin{figure}
\normalfigure{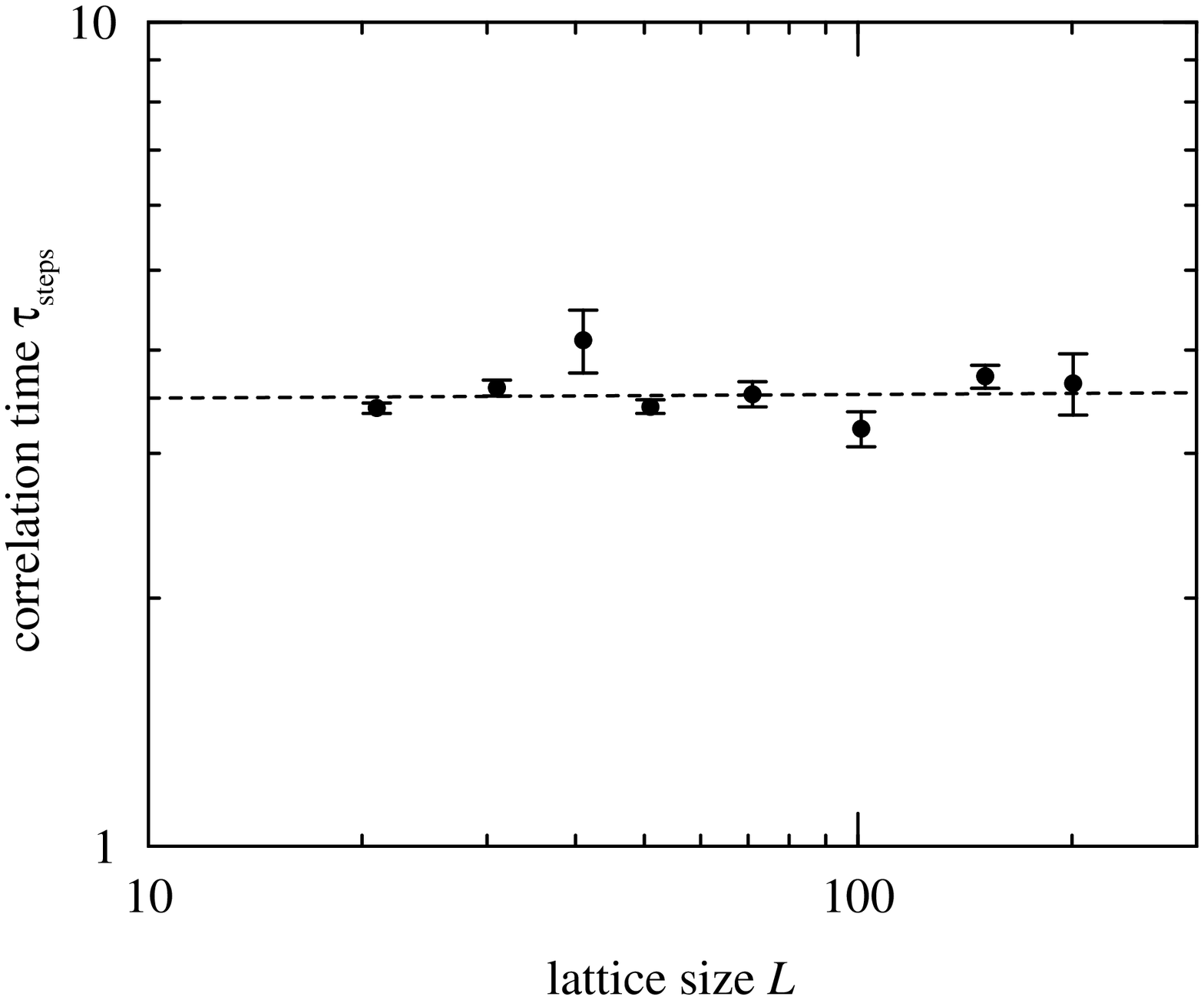}
\capt{The correlation time $\tau_{\rm steps}$ of the full-lattice
  three-colouring algorithm for the F model measured in Monte Carlo steps
  as a function of system size.  The best fit straight line gives
  $\tau_{\rm steps} \sim L^{0.005\pm0.022}$.}
\label{sw3crit}
\end{figure}

The full-lattice three-colouring algorithm does quite well at simulating
the F model, even at the critical temperature.  There is no measurable
increase in the correlation time in number of lattice sweeps with system
size at $T_c$; our best estimate of the dynamic exponent is
$z=0.005\pm0.022$.

\begin{figure}
\normalfigure{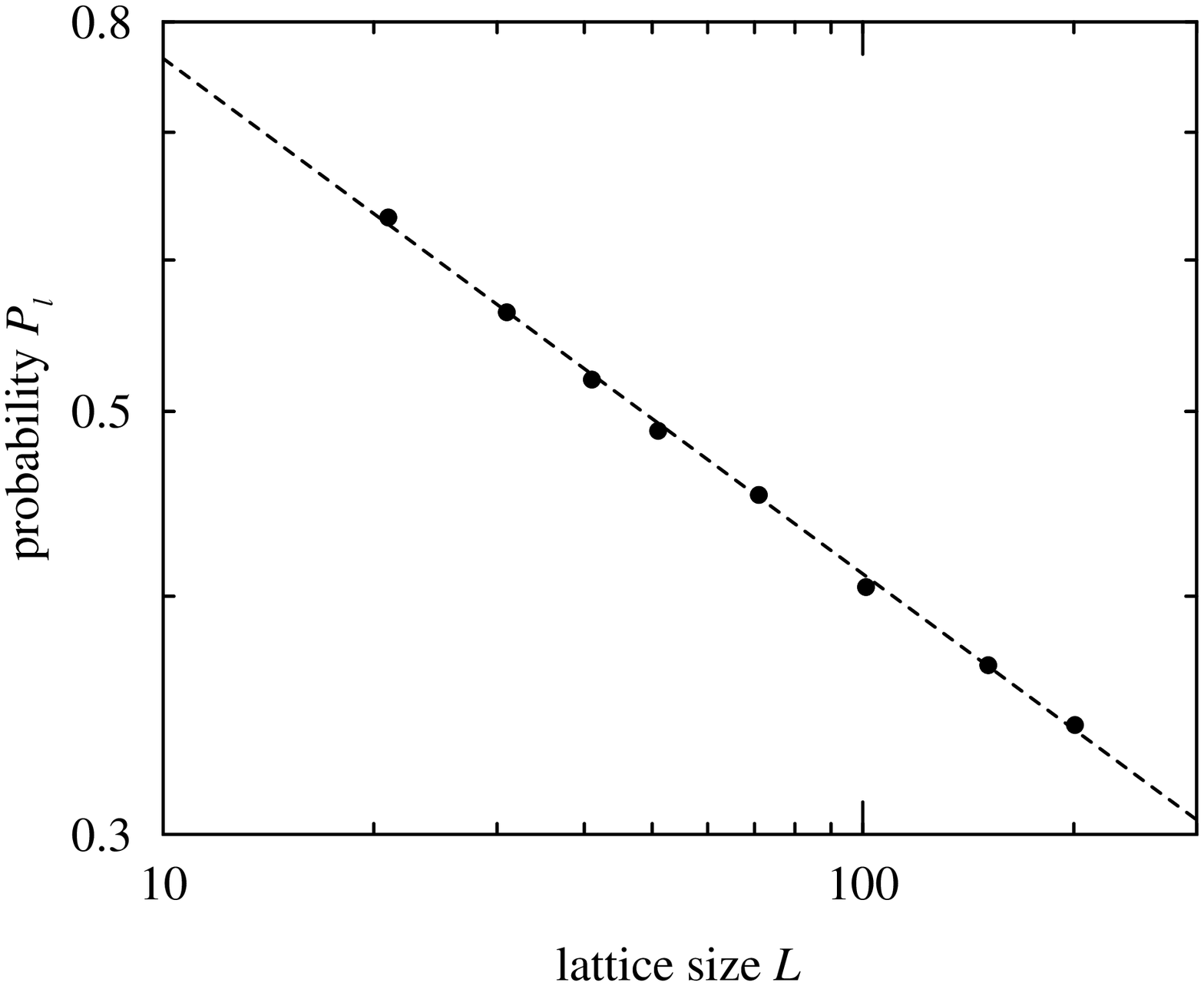}
\capt{The probability $P_l$ that a site is visited by the longest loop, as
  a function of system size $L$, for the F model at critical temperature.
  At critical temperature we find that $P_l \sim L^{-0.270\pm0.002}$, which
  is very close the exponent measured in the case of square ice
  (Section~\ref{fltcsq}).}
\label{Floop}
\end{figure}

Because of the infinite order of the phase transition in the F model we
cannot define critical exponents in the normal fashion which describe
power-law behaviour of the order parameters as we approach criticality.
However, there are a number of non-trivial exponents governing the
behaviour of the model at the critical temperature.  As noted previously,
the configurations of an ice model on a square lattice can be represented
as sets of closed loops covering the entire lattice, and the F model
corresponds to such a loop-covered system in which the loops have
``stiffness'': symmetric vertices correspond to straight segments of the
loop and are energetically favoured in the F model.  Using our full-lattice
three-colouring algorithm we have measured the probability $P_l$ that a
site is visited by the largest loop in this representation of the model,
just as we did for square ice in Section~\ref{fltcsq}.  The results are
presented in Figure~\fref{Floop}.  At the critical temperature, the data
are well fitted by a power law with an exponent of $-0.27$, very close to
the value in the square ice case, indicating that introduction of stiffness
to the loops does not significantly influence the value of this exponent.

\begin{figure}
\normalfigure{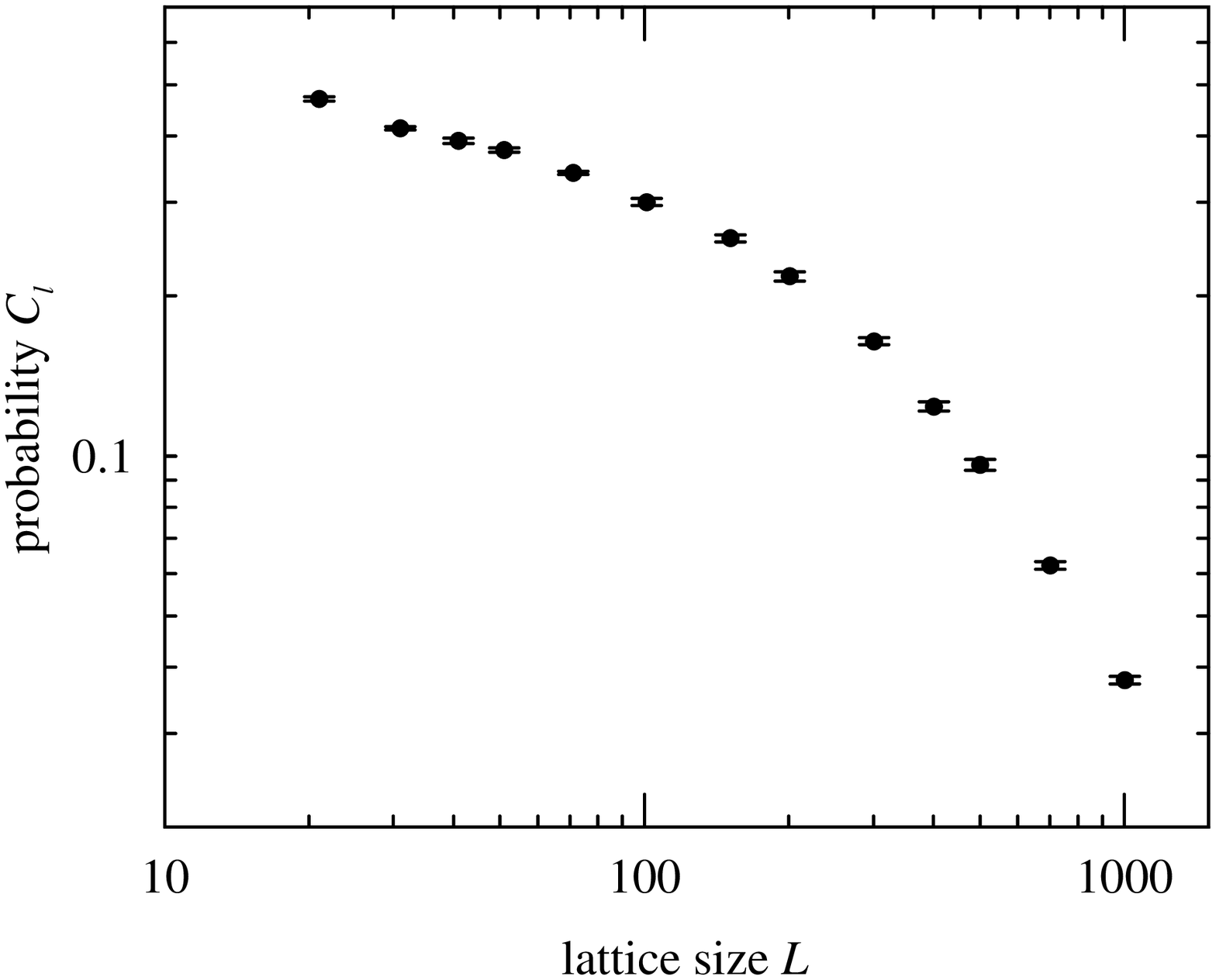}
\capt{The probability $C_l$ that a site is part of the largest cluster, as
  a function of system size $L$, for the F model at critical temperature.}
\label{Fcluster}
\end{figure}

We have also used our Monte Carlo algorithm to measure as a function of
system size $L$ the probability $C_l$ that at $T_c$ a given site is part
of the largest (percolating) cluster of nearest-neighbour symmetric
vertices.  The results are shown in Figure~\ref{Fcluster}.  Interestingly,
there is no clear power-law behaviour in these data, despite the fact that
the measurements were made at $T_c$.  Possibly this is result of strong
finite-size effects in this system.  Below the critical temperature by
contrast, the largest cluster is compact and scales as $L^2$.

\section{Conclusions}
We have described a number of Monte Carlo algorithms for simulating ice
models.  One of them, the full-lattice three-colouring algorithm, is
apparently able to simulate the F model without critical slowing down.

Using these algorithms, we have determined several exponents governing
non-local quantities in square ice and the F model.  We find that in square
ice, the average number of steps taken by a defect before it returns to its
starting point scales as $L^{1.67}$.  The probability that a site belongs
to the largest loop in the loop representation of the model scales as
$L^{-0.25}$.  In the F model, the probability of belonging to largest loop
scales with a very similar exponent $L^{-0.27}$, although the prefactor is
different.

\section*{Acknowledgements}
We are grateful to Gunter Sch\"utz for a critical reading of the
manuscript.  One of us (GTB) would like to thank the Santa Fe Institute for
their hospitality whilst this work was carried out.  This research was
funded in part by the DOE under grant number DE--FG02--9OER40542 and by the
Santa Fe Institute and DARPA under grant number ONR N00014--95--1--0975.


\begin{thebibliography}{99}
%
\bibitem{Lieb67a}
{\frenchspacing E. H. Lieb, Phys. Rev. {\bf 162}, 162 (1967).}
\bibitem{Lieb67b}
{\frenchspacing E. H. Lieb, Phys. Rev. Lett. {\bf 18}, 1046 (1967);
Phys. Rev. Lett. {\bf 19}, 108 (1967).}
%
\bibitem{Baxter82}
{\frenchspacing R. J. Baxter,
{\it Exactly Solved Models in Statistical Mechanics},
Academic Press, London, 1982.}
%
\bibitem{Rys63}
{\frenchspacing F. Rys, Helv. Phys. Acta {\bf 36}, 537 (1963).}
%
\bibitem{Bernal33}
{\frenchspacing J. D. Bernal and R. H. Fowler,
J. Chem. Phys. {\bf 1}, 515 (1933).}
%
\bibitem{Pauling35}
{\frenchspacing L. Pauling, J. Am. Chem. Soc. {\bf 57}, 2680 (1935).}
%
\bibitem{Lenard}
{\frenchspacing A. Lenard, unpublished.}
%
\bibitem{Bjerrum51}
{\frenchspacing N. Bjerrum,
K. Danske Vidensk. Selsk. Skr. {\bf 27}, 1 (1951);
Science {\bf 115}, 385 (1952).}
%
\bibitem{Rahman72}
{\frenchspacing A. Rahman and F. H. Stillinger, 
J. Chem. Phys. {\bf 57}, 4009 (1972).}
%
\bibitem{Yanagawa79}
{\frenchspacing A. Yanagawa and J. F. Nagle,
Chem. Phys. {\bf 43}, 329 (1979).}
%
\bibitem{NB96}
{\frenchspacing M. P. Nightingale and H. W. J. Bl\"ote,
Phys. Rev. Lett. {\bf76}, 4548 (1996).}
%
\bibitem{Hobbs74}
{\frenchspacing P. V. Hobbs, {\it Ice Physics,} Clarendon Press,
Oxford (1974).}
%
\bibitem{Wolff89}
{\frenchspacing U. Wolff, Phys. Rev. Lett. {\bf62}, 361 (1989).}
%
\bibitem{SW87}
{\frenchspacing R. H. Swendsen and J.-S. Wang,
Phys. Rev. Lett. {\bf58}, 86 (1987).}
%
\bibitem{Slater41}
{\frenchspacing J. C. Slater, J. Chem. Phys. {\bf 9}, 16 (1941).}
%
\bibitem{Knops79}
{\frenchspacing H. J. F. Knops, Phys. Rev. B {\bf20}, 4670 (1979).}
%
\bibitem{CMB97}
{\frenchspacing E. Carlon, G. Mazzeo, and H. van Beijeren, Phys. Rev. B
{\bf55}, 757 (1997).}
%
\end{thebibliography}
\end{document}